\documentclass[prx,twocolumn,english,superscriptaddress,floatfix,nofootinbib,aps,10pt,nolongbibliography]{revtex4-2} 
\usepackage{amsmath, amssymb, amscd, amsthm, amsfonts, amscd, dsfont, mathtools, wasysym, amsthm}
\usepackage{graphicx}%
\usepackage{dcolumn}%
\usepackage{bm}
\usepackage{standalone}
\usepackage[caption=false]{subfig}
\usepackage[utf8]{inputenc}
\usepackage{microtype}
\usepackage[plain]{algorithm}
\usepackage{algpseudocode}

\usepackage{enumitem}
\usepackage{bbm}
\usepackage{braket}
\usepackage{comment}

\usepackage{silence}
\usepackage[dvipsnames]{xcolor}
\usepackage{hyperref}
\hypersetup{
    colorlinks = true,
    linkcolor = BrickRed,%
    citecolor = PineGreen,%
    urlcolor = Blue}%

\newcommand{\ketbra}[2]{| #1 \rangle \langle #2 | }

\DeclareMathOperator{\Tr}{Tr}
\DeclareMathOperator{\Pf}{Pf}

\newcommand{\dif}{{\rm d}}
\newcommand{\iu}{{\rm i}}

\newtheorem{lemma}{Lemma}
\newtheorem{theorem}{Theorem}

\begin{document}
\title{Matchgate Circuits Deeply Thermalize}

\author{Mircea Bejan}
 \affiliation{T.C.M. Group, Cavendish Laboratory, University of Cambridge, J.J. Thomson Avenue, Cambridge, CB3 0HE, UK\looseness=-1}%
\author{Benjamin B\'eri}
\affiliation{T.C.M. Group, Cavendish Laboratory, University of Cambridge, J.J. Thomson Avenue, Cambridge, CB3 0HE, UK\looseness=-1}%
\affiliation{DAMTP, University of Cambridge, Wilberforce Road, Cambridge, CB3 0WA, UK}%
\author{Max McGinley}%
\affiliation{T.C.M. Group, Cavendish Laboratory, University of Cambridge, J.J. Thomson Avenue, Cambridge, CB3 0HE, UK\looseness=-1}%

\begin{abstract}
    We study the ensemble of states generated by performing projective measurements on the output of a random matchgate (or free-fermionic) quantum circuit. We rigorously show that this `projected ensemble'  exhibits deep thermalization: For large system sizes, it converges towards a universal ensemble that is uniform over the manifold of Gaussian fermionic states. %
    As well as proving moment-wise convergence of these ensembles, we demonstrate that the full distribution of any physical observable in the projected ensemble is close to its universal form in Wasserstein-1 distance, which we argue is an appropriate and efficiently computable measure of convergence when studying deep thermalization. Using this metric, we also numerically find that local matchgate circuits deeply thermalize after a timescale $t \sim L^2$  set by the diffusive spreading of quantum information. Our work opens up new avenues to experimentally accessible protocols to probe the emergence of quantum statistical mechanics and benchmark quantum simulators.
\end{abstract}

\maketitle

Our
theoretical understanding of how thermalization and statistical mechanics emerge from underlying microscopic many-body dynamics has evolved greatly over the last century, particularly in the context of isolated quantum systems \cite{DeutschRMT, SrednickiRMT, Kaufman2016, Neill2016}. A standard approach to probing thermalization is to watch how small subsystems $A$ locally equilibrate under the full dynamics of the system. At late times, statistical mechanics prescribes that the state of $A$ should have maximal entropy (subject to any physical constraints). Microscopically, this is made possible by the generation of entanglement between $A$ and its surroundings $B$ \cite{Popescu2006,Goldstein2006}, the latter acting as a heat bath.

Thanks to advances in quantum simulators \cite{EhudPRXQroadmap}, it is now possible to experimentally study quantum dynamics beyond the evolution of local observables. This has motivated notions of generalized forms of thermalization \cite{Cotler2023, JChoi_2023, HoChoi2022,  Ippoliti2022solvablemodelofdeep, Claeys2022, MaxMichelePRL, Tanmay2023, Ippoliti2023purification, dgge, shrotriya2023nonlocalitydeepthermalization, vairogs2024extractingrandomnessquantummagic, mark2024maximumentropyprincipledeep, liu2024deepthermalizationcontinuousvariablequantum,  chang2024deepthermalizationchargeconservingquantum, Varikuti2024unravelingemergence}, which pertain to 
full conditional distributions on $A$ given  measurement outcomes on $B$, rather than ignoring $B$ entirely.
Specifically, if one looks at the statistical properties of the ensemble of \emph{states} on $A$ after measurements on $B$, one gets the so-called \textit{projected ensemble} (PE)~\cite{Cotler2023}, see also Refs.~\onlinecite{pe1, pe2, pe3, pe4, pe5, pe6}.
This contains more information than traditional measures of thermalization. Results have pointed to the emergence, at late times, of a new phenomenon dubbed \textit{deep thermalization} (DT)~\cite{HoChoi2022, Ippoliti2022solvablemodelofdeep}, where the PE approaches a universal ensemble of states that are
maximally random, up to physical constraints.
Probing deep thermalization experimentally comes with a postselection problem, akin to measurement-induced phase transitions in entanglement~\cite{li2018zeno, li2019mipt, choi2020qecc, ippoliti2021mom, fisher2023rqc, skinner2019mipt, versini2024}. Since the PE contains $\sim 2^L$ states (for an $L$-qubit system), brute-force tests of DT have exponential sample complexity and have thus been restricted to dozens of qubits~\cite{Cotler2023, JChoi_2023}.
At the same time, the only rigorous theoretical results on DT in finite-dimensional Hilbert spaces are for Haar-random qubit states~\cite{Cotler2023, MaxMichelePRL} and dual-unitary circuits~\cite{HoChoi2022, Claeys2022}, for which no known solution to the postselection problem exists.
This raises the question of whether one can experimentally realize systems that exhibit deep thermalization, but which do not suffer from a postselection barrier.

\begin{figure}[t]
    \centering
        \includegraphics[width=8.6cm]{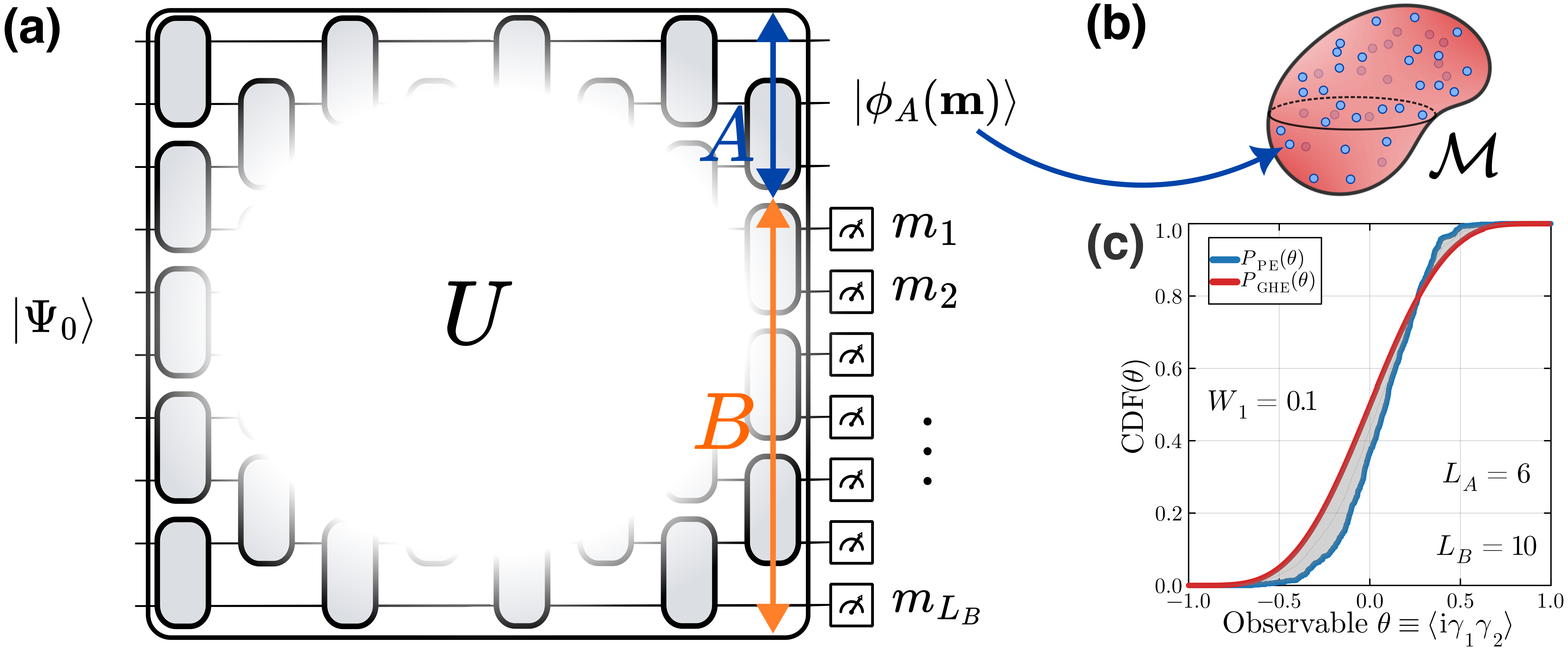}
    \caption{%
             (a) Starting from an initial state $\ket{\Psi_0}$  on $L$ fermionic modes, a Gaussian circuit $U$ is applied, after which Fock basis measurements are performed on subsystem $B$, with outcome $\mathbf{m}$. This yields pure states $\ket{\phi_A (\mathbf{m})}$, which constitute the PE.
            (b) The PE is a discrete ($2^{L_B}$ elements shown as blue circles) probability distribution on the manifold $\mathcal{M}$ of pure Gaussian states. We prove that the PE covers $\mathcal{M}$ uniformly for $L_B \to \infty$.
            (c) Cumulative distribution function for the projected and Haar ensembles vs the observable $\theta = \langle \iu \gamma_1 \gamma_2 \rangle$
            for global~$U$. The Wasserstein-1 distance $W_1$ (grey shaded region's area) quantifies the distance between these distributions. %
            }
    \label{fig:setup}
\end{figure}

In this work, we study DT in free-fermion dynamics and their qubit-based equivalents, matchgate circuits~\cite{matchgates1valiant,  matchgates3jozsa, matchgates2terhalDV}, cf.~Fig.~\ref{fig:setup}. These circuits, which form the basis of fermion linear optics (FLO)~\cite{matchgates2terhalDV, bravyi2004gaussian}, can capture the time evolution of any quadratic fermionic Hamiltonian.
Because matchgates are classically efficiently simulable, one can construct hybrid quantum-classical algorithms for experimentally probing DT in matchgate-dominated systems, which overcome the postselection barrier~\cite{psf1, psf2, psf3}. Accordingly, the physics we study here can be efficiently verified and may serve as a robust benchmark for quantum simulators and computers.
While these circuits are not universal for quantum computation~\cite{fqc2002} and hence cannot exhibit exactly the same behavior as fully universal systems, here we prove that free fermions do deeply thermalize: In particular, at late times, the projected ensemble tends towards a universal distribution, which we refer to as the Gaussian Haar ensemble (GHE). This rigorously establishes that the phenomenology seen in the numerical experiments of Ref.~\onlinecite{dgge} holds true in general and complements a related observation made for bosonic Gaussian systems in Ref.~\onlinecite{liu2024deepthermalizationcontinuousvariablequantum}.

Our main rigorous results---Theorems \ref{thm:Moments}, \ref{thm:MomentsManyBody}, and \ref{thm:Wasserstein}---show that, for random free-fermionic circuits, the PE converges to the GHE at the level of \textit{full probability distributions} of physical observables, which is much stronger than previous approaches where low-order moments are compared.
To make this comparison concrete, we propose to use a particular measure of closeness between probability distributions known as the Wasserstein-1 distance,  which we argue is particularly relevant to this setting.
We supplement these results with numerics, which also show that the corresponding probability distributions remain close in the same precise sense also for local random matchgate circuits at late times. We argue that the universal behavior emerges after a timescale $t \sim L^2$ set by the diffusive spreading of quantum information in the system.
\textit{\textbf{Physical setup and the projected ensemble.}---} 
We consider
Gaussian dynamics~\cite{matchgates2terhalDV, bravyi2004gaussian} (see also~\cite{SM}) 
%
%
%
on $L$ fermionic modes, which are spanned by $2L$ Majorana operators $\{\gamma_j\}$ satisfying $\{\gamma_j, \gamma_k\} = 2\delta_{jk}$~\cite{Kitaev_unpaired_2001, fqc2002}. Starting from the vacuum $\ket{0^{\otimes L}}$ without loss of generality, we apply a sequence of FLO gates, without particle number conservation~\cite{matchgates2terhalDV, bravyi2004gaussian}. These operations can be specified by an orthogonal matrix $O \in \textrm{SO}(2L)$, to which we associate a Gaussian unitary operator $U[O]$ in Fock space~\cite{Kitaev_unpaired_2001} (or, alternatively, a matchgate circuit on $L$ qubits~\cite{matchgates2terhalDV}). Often, we will consider a (1+1)D geometry with gates applied in a brickwork architecture as shown in Fig.~\ref{fig:setup}(a). Being Gaussian, the resulting state $\ket{\Psi} = U[O]\ket{0^{\otimes L}}$ is fully determined by its covariance matrix $\Gamma_{jk} \coloneqq \frac{\iu}{2} \braket{\Psi|[\gamma_j, \gamma_k]|\Psi}$~\cite{fqc2002}.

We then projectively measure each mode within a subsystem $B \subset [L] \coloneq \{1, \dots, L\}$ of size $L_B = |B|$ in the Fock basis. The outcome $\mathbf{m} \in \{ 0,1 \}^{L_B}$ occurs with probability $p_\mathbf{m} = \bra{\Psi} (\mathbbm{1}_A \otimes \ketbra{\mathbf{m}}{\mathbf{m}}_B) \ket{\Psi}$, conditioned on which the post-measurement state of the unmeasured modes $A = [L] \backslash B$ becomes $\ket{\phi_A(\mathbf{m})} = (\mathbbm{1}_A \otimes \bra{\mathbf{m}}_B) \ket{\Psi} / \sqrt{p_\mathbf{m}}$, which is also a Gaussian state~\cite{matchgates2terhalDV,fqc2002,bravyi2004gaussian}, and hence is specified by a covariance matrix $\Gamma^A_{\mathbf{m}}$. The distribution of post-measurement states is the PE [see Fig.~\ref{fig:setup}(b)]
\begin{align}
    \mathcal{E}_\mathrm{PE} \coloneq \{ p_\mathbf{m}, \ket{\phi_A(\mathbf{m})} \}.
    \label{eq:PEDef}
\end{align}%

Our aim is to characterize universal properties of the PE for $L_B \to \infty$. Recent work showed that for generic dynamics, the PE tends towards the Haar ensemble, i.e., the uniform distribution over all states on $A$ \cite{Cotler2023, HoChoi2022, Claeys2022, Ippoliti2022solvablemodelofdeep, MaxMichelePRL, Tanmay2023, Ippoliti2023purification, shrotriya2023nonlocalitydeepthermalization}. More generally, if the dynamics is constrained, e.g., for energy- or charge-conserving dynamics, or in Gaussian bosonic systems, it has been suggested that the appropriate ensemble is the one that is maximally entropic subject to these constraints \cite{mark2024maximumentropyprincipledeep, liu2024deepthermalizationcontinuousvariablequantum, chang2024deepthermalizationchargeconservingquantum}. 

For free-fermions, the authors of Ref.~\onlinecite{dgge} presented numerical evidence, in terms of certain statistical moments,
that under Hamiltonian evolution a generalized Gibbs ensemble of a similar kind is reached at late times.
However, analytical results on finite-order statistical moments and, especially, on the full distribution of post-measurement states have been lacking.
Here, we rigorously prove that the PE does indeed converge (in a sense that we will make precise) to the \textit{Gaussian Haar ensemble},
which is uniform over all Gaussian states on $A$
\begin{align}
    \mathcal{E}_\mathrm{GHE} \coloneq \text{Unif}\Big( \big\{ \ket{\phi^A} \in \mathcal{H}^A : \ket{\phi^A} \textrm{ is Gaussian}\} \Big).
    \label{eq:GHEDef}
\end{align}
Since we can identify Gaussian states, through their covariance matrix, 
with points on the symmetric space $\frac{\mathrm{SO}(2L_A)}{\mathrm{U}(L_A)}$, or DIII in Cartan's classification \cite{AltlandZirnbauer} (see~\cite{SM}), `$\mathrm{Unif}$' here refers to the corresponding circular ensemble~\cite{dahlhaus_prb_2010}.
\textit{\textbf{Deep thermalization via moments.}---}
Because the post-measurement states are Gaussian, a natural way to characterize the PE \eqref{eq:PEDef} is via the covariance matrices of the conditional states $\Gamma^A_{\mathbf{m}}$. Viewing 
the matrix elements 
$[\Gamma^A_{\mathbf{m}}]_{ij}$ as a collection of $(2L_A)^2$ random variables, we define
the $k$th moments of this multivariate distribution
\begin{align}
    \mathcal{C}^{(k)}_{\rm PE, \mathbf{i} \mathbf{j}} = \sum_{\mathbf{m}} p_{\mathbf{m}} [\Gamma^A_\mathbf{m}]_{i_1j_1} \cdots [\Gamma^A_\mathbf{m}]_{i_kj_k},
    \label{eq:MomentMatrixElement}
\end{align}
where $\mathbf{i} = (i_1, \ldots, i_k)$ is a multi-index where each individual index $i_a \in [2L_A]$ specifies a Majorana operator $\gamma_{i_a}$ (similar for $\mathbf{j}$). To probe deep thermalization, we compare $\mathcal{C}^{(k)}_{\mathrm{PE}, \mathbf{i} \mathbf{j}} $ with the corresponding GHE moments $\mathcal{C}^{(k)}_{\mathrm{GHE}, \mathbf{i} \mathbf{j}}$.
Our first rigorous result (proved in~\cite{SM}) is that free fermions evolving under globally random Gaussian circuits deeply thermalize at the level of these moments.
\begin{theorem} \label{thm:Moments}
    If $\ket{\Psi}$ is prepared by a circuit $U[O]$ 
    with $O$ drawn from the Haar measure over $\textup{SO}(2L)$, then for any choice of $\mathbf{i}$, $\mathbf{j}$, the probability that the moment \eqref{eq:MomentMatrixElement} deviates by more than $\epsilon$ from the corresponding GHE moment is
    \begin{align}
        \mathbbm{P}_{O}\left( |\mathcal{C}^{(k)}_{\textup{PE}, \mathbf{i} \mathbf{j}} - \mathcal{C}^{(k)}_{\textup{GHE}, \mathbf{i} \mathbf{j}}| > \epsilon \right) < \exp\left( -\frac{\kappa L_B \epsilon^2}{k^2}\right)
        \label{eq:MomentConcentrate}
    \end{align}
    for some constant $\kappa \geq 1/64$.
\end{theorem}
\noindent This result, observed numerically for $k \leq 4$
in Ref.~\onlinecite{dgge}, establishes that the moments of the covariance matrices themselves are close to those of the GHE with high probability in the large $L_B$ regime. 

We may also consider a more operational notion of deep thermalization, which is more closely related to the non-Gaussian case. Here, we compare the moments of the full many-body states
\begin{align}
    \mathcal{E}^{(k)}_{\textrm{PE}} \coloneqq  \sum_\mathbf{m} p_\mathbf{m} \big(\ket{\phi_A(\mathbf{m})}\bra{\phi_A(\mathbf{m})}\big)^{\otimes k}
    \label{eq:MomentManyBody}
\end{align}
 with those of the GHE, $\mathcal{E}^{(k)}_{\textrm{GHE}}$. Following the nomenclature for fully random quantum states, we say that the projected ensemble forms a \textit{$\epsilon$-approximate Gaussian $k$-design} if %
\begin{align}
\label{eq:DeltaKDef}
    \Delta^{(k)} \coloneq \left\|\mathcal{E}^{(k)}_{\textrm{PE}} - \mathcal{E}^{(k)}_{\textrm{GHE}} \right\|_1 \leq \epsilon
\end{align}
where $\|X\|_1 \coloneqq \Tr[\sqrt{X^\dagger X}]$ is the trace norm. Concretely, this condition implies that any strategy to distinguish these ensembles using $k$ copies of the state---including those using non-Gaussian operations---has success probability close to that of a random guess $p_{\rm succ} \leq \frac{1}{2} + \frac{\epsilon}{4}$ \cite{Helstrom1969}.
We can prove the following. (See~\cite{SM} for a proof.)
\begin{theorem} \label{thm:MomentsManyBody}
    There exist constants $\alpha$, $\beta$ such that if $\ket{\Psi}$ is prepared by a circuit $U[O]$ 
    with $O$ drawn from the Haar measure over $\textup{SO}(2L)$, then the PE forms an $\epsilon$-approximate Gaussian $k$-design with probability at least
    $1 - \delta$, provided $L_B \geq \alpha \log(1/\delta)/\epsilon^2$ and
    \begin{align}
        kL_A \leq \beta\log(L_B).
    \end{align}
\end{theorem}
\noindent This should be compared to the case with $\ket{\Psi}$ fully Haar-random, where the PE forms an $\epsilon$-approximate $k$-design for $kL_A \leq \beta' L_B$, provided $L_B \geq \alpha'[\log(1/\epsilon) + \log \log (1/\delta)]$ for some constants $\alpha',\beta'$ \cite{Cotler2023}. 

An important concept that we invoke in our proofs of Theorems~\ref{thm:Moments} and~\ref{thm:MomentsManyBody} is the concentration of measure phenomenon for the Haar measure over $\mathrm{SO}(2L) \ni O$~\cite{Anderson2009}, which states that for certain well-behaved functions $F(O)$, large deviations from the mean are very rare. In brief, we show that the quantities~(\ref{eq:MomentMatrixElement}, \ref{eq:MomentManyBody}) satisfy the necessary requirements to invoke these results, and that their mean is equal to their GHE equivalent.%

\textit{\textbf{Deep thermalization via Wasserstein distance.}---} 
The $k$th moments of the covariance matrices \eqref{eq:MomentMatrixElement} and the many-body states \eqref{eq:MomentManyBody} allow us to infer certain statistics of a number of physical quantities in the projected ensemble: For instance, the average and variance of the purity of a subsystem of $A$ can be computed from $\mathcal{E}^{(4)}_{\rm PE}$. However, it would be desirable to characterize the full distribution of observables in the PE more comprehensively, without having to look at a spectrum of distances $\Delta^{(k)}$ for all $k$. For this reason, we look to relate the ensembles $\mathcal{E}_\mathrm{PE}$~\eqref{eq:PEDef} and $\mathcal{E}_\mathrm{GHE}$~\eqref{eq:GHEDef} using a \textit{statistical distance}.

Statistical
distances allow one to compare two probability distributions~\cite{Cover2005}. These include the total variation distance and the Kullback-Leibler divergence~\cite{kullback1959}, which are natural choices in many probability-theoretic settings. However, these particular examples always evaluate to their maximal
value whenever we compare a discrete with a continuous distribution, even if the former is a `good' discretization of the latter; this is discussed in~\cite{SM}.

Here, instead, we propose to use the \textit{Wasserstein-1 distance}~\cite{Kantorovich1942,Villani2009} to
quantify deep thermalization. Formally, for two distributions $\mu$, $\nu$ over a manifold $\mathcal{M} \ni x$, %
the Wasserstein-1 distance is
\begin{align}
    W_1(\mu, \nu) = \sup_{f(x) : \text{1-Lipschitz}} \mathbbm{E}_{x \sim \mu}[f(x)] - \mathbbm{E}_{x \sim \nu}[f(x)],
    \label{eq:WassersteinDefMain}
\end{align}
where $\mathbbm{E}_{x \sim \mu} [f(x)]$ denotes the expectation of $f$ over the distribution $\mu$, and similar for $\nu$.
Here, we are using the `test function' $f(x)$ to probe discrepancies between the distributions $\mu$ and $\nu$. In particular, we take the maximum discrepancy over all `sufficiently smooth' functions, specifically those that are $\eta$-Lipschitz with $\eta \leq 1$, 
\begin{align}
    |f(x_1) - f(x_2)| &\leq \eta\, d(x_1, x_2) & \forall x_1, x_2 \in \mathcal{M} %
\end{align}
where $d(\cdot, \cdot)$ is a chosen metric on $\mathcal{M}$. In our setting, we are comparing distributions $\mu=\mathcal{E}_\mathrm{PE}$~\eqref{eq:PEDef} and $\nu=\mathcal{E}_\mathrm{GHE}$~\eqref{eq:GHEDef}, with $\mathcal{M}$ being the manifold
of pure Gaussian states. As we shall see, $1$-Lipschitz functions over this manifold
encompass physically relevant properties~\cite{SM}.

Computing the Wasserstein-1 distance of $\mathcal{E}_\mathrm{PE}$  and $\mathcal{E}_\mathrm{GHE}$ is a formidable task since we need to optimize over functions $f$. However, if we instead look at the distribution of a particular scalar quantity $\theta = \theta(\Gamma^A)$, we can make some progress. In this case, we need to compare two univariate distributions, which can be represented by their density functions $p(\theta)$ and  $q(\theta)$ over some domain $X \subseteq \mathbbm{R}$. Eq.~\eqref{eq:WassersteinDefMain} then simplifies significantly to~\cite{w1CDFs}
\begin{align}
    W_1\big(p(\theta),q(\theta)\big) = \int_{\theta \in X} \dif \theta |P(\theta) - Q(\theta)|,
    \label{eq:W11D}
\end{align}
with $P(\theta)$ and $Q(\theta)$ being the cumulative distribution functions (CDFs) of $p(\theta)$ and $q(\theta)$, respectively, i.e.,~one takes the unsigned area between the CDFs [see Fig.~\ref{fig:setup}(c)]. This will allow us to numerically compute $W_1$ in certain instances later on.%

\begin{figure*}[t]
    \centering
        \includegraphics[width=17.8cm]{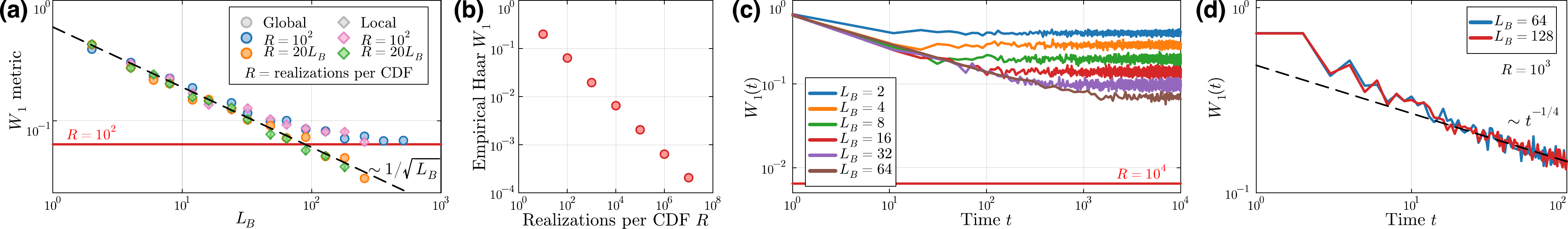}
    \caption{%
             Deep thermalization of fermionic Gaussian dynamics: Wasserstein-1 distance $W_1$ between distributions of the correlator $\theta(\Gamma^A_{\mathbf{m}}) =\braket{\iu \gamma_1\gamma_2}_{\mathbf{m}}$ for $L_A=2$, averaged over $N=10^2$ circuit instances.
             (a) $W_1$ decays with the number of measured qubits $L_B$ as $W_1 \propto 1/\sqrt{L_B}$ (dashed line), for both global (circles) and late-time local ($t=10^4$, rhombi) dynamics, down to a threshold value 
             (horizontal red line).             
             The error bars, where visible, show the standard error of the mean (SE).
             (b) Estimate of the finite-$R$ plateau from comparing an empirical CDF using $R$ GHE realizations and the exact GHE CDF, showing $W_1 \propto 1/\sqrt{R}$.
             (c) $W_1(t)$ vs time $t$ for local dynamics. The horizontal red line shows the empirical threshold. Relative SEs are all below $9\%$ (not shown).
             An empirical PE is constructed every 10 ($10^2$) time steps for $L_B = 2, \dots, 32$ ($64$) and $R=10^4$.
             (d) Closer view of the decay $W_1(t) \propto t^{-1/4}$ (dashed line) for $R=10^3$ and an empirical PE built at each time step. Relative SEs are all below $9\%$.
            }
    \label{fig:W1_fig}
\end{figure*}

Our final rigorous result
uses the Wasserstein-1 distance to characterize the \textit{full distribution} of any (sufficiently well-behaved) physical quantity $\theta$, beyond low-order moments.
\begin{theorem}\label{thm:Wasserstein}
    Let $\theta(\Gamma^A)$ be a physical quantity that is bounded in an interval $[a,b]$, differentiable, and $c$-Lipschitz, i.e., $\|\nabla_\Gamma \theta\|_2 \leq c$. If $\ket{\Psi}$ is prepared by a circuit $U[O]$ 
    with $O$ drawn from the Haar measure over $\textup{SO}(2L)$,
    then the distributions of $\theta$ in the projected and Gaussian Haar ensembles $p_{\rm PE}(\theta)$ and $p_{\rm GHE}(\theta)$ converge in Wasserstein-1 distance for large $L_B$ %
    \begin{align}
        W_1\left(p_{\rm PE}(\theta), p_{\rm GHE}(\theta)\right) \leq \kappa \frac{\sqrt{\log L_B}}{L_B^{1/6}},
        \label{eq:W1Decay}
    \end{align}
    for some constant $\kappa \leq 27\max(c,b-a)$.
\end{theorem}

Our proof,
given in~\cite{SM}, uses finite-degree polynomial approximations of Lipschitz functions~\cite{Jackson1912}, combined with aforementioned concepts from concentration of measure~\cite{Anderson2009}.
Theorem~\ref{thm:Wasserstein} implies %
that, for large~$L_B$, a full histogram of $\theta$ in the PE will be very close to the corresponding histogram in the GHE
(provided the bins are not too narrow) \cite{NilesWeed2022}. Physically relevant functions $\theta$ that satisfy the criteria above include $n$-point correlation functions $\braket{\gamma_{j_1} \cdots \gamma_{j_n}}$, subsystem R\'enyi entropies, and von Neumann entanglement entropies, the latter of which is not directly related to any integer moment $\mathcal{E}^{(k)}_{\rm PE}$.

%

%

%

%

%

%

%

%
\textit{\textbf{Numerics.}---} %
Our rigorous results support the claim that a state $\ket{\Psi}$ prepared by a random Gaussian unitary will typically exhibit deep thermalization with respect to $\mathcal{E}_\mathrm{GHE}$~\eqref{eq:GHEDef}. Here, we supplement these theorems with numerical results, both for globally and locally random Gaussian (1+1)D circuits. Thanks to~\eqref{eq:W11D}, we can evaluate the Wasserstein-1 distance for the distributions of physical quantities $\theta(\Gamma^A_\mathbf{m})$ efficiently (i.e.,~without having to consider all $2^{L_B}$ possible outcomes), as follows. We first get an estimate of the CDF $P_{\rm PE}(\theta)$ by generating measurement outcomes $\mathbf{m}^{(i)}$ ($i=1,\ldots,R$) on $B$~\cite{SM}, computing $\Gamma^A_{\mathbf{m}^i}$ and from these $\theta^{(i)} = \theta(\Gamma^A_{\mathbf{m}^i})$.  This determines an empirical CDF $\hat{P}_{\rm PE}(\theta) = \frac{1}{R}\sum_{i} \Theta(\theta^{(i)}-\theta)$, which can be combined with analytically known or empirically estimated forms for $P_{\rm GHE}$ to get an estimate $W_1(\hat{P}_{\rm PE},P_{\rm GHE})$. We then average this over $N$ circuit instances to estimate $\mathbbm{E}_\Psi W_1$. %

\textit{Global dynamics.---} 
First, we take global Gaussian random unitaries, and look at the distribution of a particular correlation function $\theta = \braket{\iu \gamma_1 \gamma_2}_\mathbf{m} = [\Gamma^A_\mathbf{m}]_{12}$, using analytically known expressions for $P_\mathrm{GHE}(\theta)$~\cite{SM}. Taking $L_A = 2$, our results agree well with a power-law decay $W_1 \propto 1/\sqrt{L_B}$. When $R$ is finite, this saturates at large $L_B$ due to the finite sample-size [Fig.~\ref{fig:W1_fig}(a), (b)]. Similar results hold for other values of $L_A$~\cite{SM}.
This behavior is consistent with the bound $W_1 \lesssim L^{-1/6}_B$ from Thm.~\ref{thm:Wasserstein}, but does not saturate it; thus, it may be possible to quantitatively improve~\eqref{eq:W1Decay}.
By analogy to Ref.~\onlinecite{Cotler2023}, an estimate of the 
plateau value can be obtained by considering an empirical CDF constructed from $R$ GHE realizations; its discrepancy is shown as the horizontal line in Fig.~\ref{fig:W1_fig}(a), and plotted in Fig.~\ref{fig:W1_fig}(b). Taking $R$ to scale linearly $L_B$ alleviates this numerical artifact, see End Matter.

%
%
%
%

%

%
%

%

%

%
\textit{Local dynamics.---} We now take local (1+1)D circuits, with each gate a random FLO unitary, which can be implemented by matchgates \cite{matchgates1valiant, matchgates2terhalDV, matchgates3jozsa}. Since any Gaussian unitary can be implemented by a linear-depth (1+1)D circuit \cite{floU1, floU2}, we anticipate that random local Gaussian circuits will converge towards globally random Gaussian unitaries reasonably quickly; thus, we expect DT at long timescales. Indeed, our simulations suggest a power-law decay $W_1 (t) \propto t^{-1/4}$ [Fig.~\ref{fig:W1_fig}(c), (d)] down to a late-time value set by the size of the measured subsystem $L_B$ following $W_1(t\gg 1)\propto 1/\sqrt{L_B}$ [Fig.~\ref{fig:W1_fig}(a), (c)].
Here, we choose $R$ large enough such that the empirical CDF $\hat{P}_\mathrm{PE}$ approximates well $P_\mathrm{PE}$, see End Matter. 
%
%
%
We also verified this $W_1 (t) \propto t^{-1/4}$ behavior for other choices of $L_A$~\cite{SM}. 

\textit{From local to global dynamics via diffusion.---} 
How fast information spreads under local dynamics sets the time-dependence of $W_1(t)$, and we can relate $W_1(t) \propto t^{-1/4}$ for local dynamics to the observed $W_1 \propto 1/\sqrt{L_B}$ for global dynamics. 
Since the dynamics does not conserve particle number, this spreading is not associated with any conserved charge. Instead, it simply quantifies the spreading of correlations. We support diffusive spreading with numerical evidence in~\cite{SM}.
%
%
Diffusive spreading implies that at time $t$, the effective number of measured modes influencing the PE is $\ell \sim \sqrt{t}$. 
By $W_1 \propto 1/\sqrt{L_B}$ for $L_B$ measured modes for global dynamics, we expect $W_1(t) \propto 1/\sqrt{\ell} \sim t^{-1/4}$ for local dynamics, which agrees with our simulations [Fig.~\ref{fig:W1_fig}(d)]. 

\begin{figure}[h]%
    \centering
        \includegraphics[width=8.6cm]{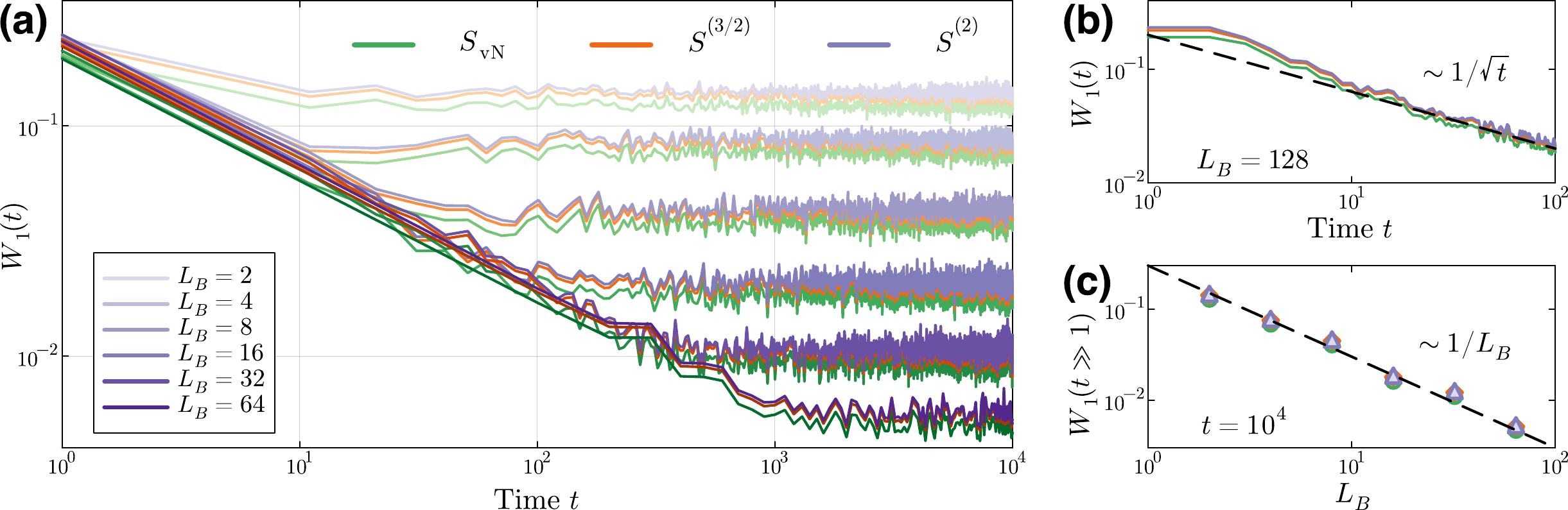}
    \caption{%
            Deep thermalization of local fermionic Gaussian dynamics: Wasserstein-1 distance $W_1$ between distributions of subsystem von Neumann or R\'enyi entropy $\theta(\Gamma^A_{\mathbf{m}}) = S^{(\alpha)}(\rho^{A_1}_\mathbf{m})$ for $L_{A_1}=\frac{L_A}{2}=1$,  $N=10^2$ circuit instances.
             (a) $W_1(t)$ saturates as $\propto 1/\sqrt{t}$, down to a value set by the measured subsystem size $L_B$. Here, $R=10^4$, and relative errors are all below $12\%$.
             (b) Closer view of the decay $W_1(t) \propto 1/\sqrt{t}$ (dashed line) for $R=10^3$.
             (c) Scaling of $W_1 (t\gg 1) \propto 1/L_B$ (dashed line), imperceptible error bars~(SE). 
            }
    \label{fig:W1_entropies}
\end{figure}

\textit{Entanglement entropies.---} %
To further test DT,
we consider the distribution of subsystem von Neumann and R\'enyi entanglement entropies $\theta = S_\mathrm{vN}, S^{(\alpha)}$. Since we cannot access the analytical form for $P_\mathrm{GHE}(\theta)$, we build an empirical $\hat{P}_\mathrm{GHE}(\theta)$ from $10^8$ realizations. Taking $L_A = 2$, $L_{A_1}=L_A/2$ and computing $S^{(\alpha)}(\rho^{A_1}_\mathbf{m})$ for $\alpha \in \{ 1, \frac{3}{2}, 2\}$, our numerics suggest a power-law decay $W_1(t)\propto 1/\sqrt{t}$ [Fig.~\ref{fig:W1_entropies}(a), (b)], approaching a late-time value $W_1(t\gg 1 ) \propto 1/L_B$ [Fig.~\ref{fig:W1_entropies}(c)], for all $\alpha \geq 1$. We can again understand the relaxation from local to global dynamics behavior of $W_1$ through the diffusion argument $W_1(t) \propto 1/\ell \sim 1/\sqrt{t}$. 
The late-time decay of $W_1$ for $\alpha \in \{1, \frac{3}{2}, 2 \}$ is consistent with the bound $W_1 \lesssim L^{-1/6}_B$ from Thm.~\ref{thm:Wasserstein}, but does not saturate it.

\textit{\textbf{Discussion and outlook.}---} %
We have rigorously shown that random free-fermion states deeply thermalize. Specifically, we showed that the projected ensemble tends towards a particular universal distribution---the Gaussian Haar ensemble---at the level of both statistical moments (in keeping with previous methodology \cite{Cotler2023,HoChoi2022,MaxMichelePRL,Claeys2022, Ippoliti2022solvablemodelofdeep}), but also using the more general Wasserstein-1 distance $W_1$. %
Our rigorous theorems are complemented by numerical simulations, which also show a power-law convergence with depth $t$ for local circuits, which we attribute to the diffusive spreading of quantum information. %
A key distinction between Gaussian and universal dynamics seems to be that the former typically shows power-law decay of $W_1$ and moments, while the latter shows exponential decay, with respect to $L_B$ or~$t$.

Because matchgate circuits are efficiently simulable,
these phenomena can be experimentally probed in a scalable manner, using classical side-processing to avoid postselection issues \cite{psf2,psf3}. This form of DT, therefore, offers a versatile way to benchmark quantum simulators. While similar approaches might apply to Clifford circuits as well, we find that the PE there exhibits a much more trivial structure, forming at most a 1-design~\cite{SM}. Moving forward, it would be interesting to study how these phenomena are enriched upon doping with a small number of non-Gaussian gates. 
In addition, using this formalism to identify observables whose distributions show a stark contrast between Gaussian and universal dynamics could help study phase transitions between classically efficiently simulable and hard-to-simulate regimes of quantum dynamics. Finally, expanding our analysis to account for realistic experimental imperfections by including the effects of noise and measurement errors would be appealing. %

\textit{
\textbf{Acknowledgments.}---} We thank Jan Behrends and Wen Wei Ho for useful discussions. 
This work was supported by EPSRC PhD Studentship 2606484, EPSRC grant EP/V062654/1, and by Trinity College, Cambridge.
Our simulations used resources at the Cambridge Service for Data Driven Discovery operated by the University of Cambridge Research Computing Service (\href{www.csd3.cam.ac.uk}{www.csd3.cam.ac.uk}), provided by Dell EMC and Intel using EPSRC Tier-2 funding via grant EP/T022159/1, and STFC DiRAC funding  (\href{www.dirac.ac.uk}{www.dirac.ac.uk}).

\begin{center}
    \textbf{{\fontsize{11}{11}\selectfont End Matter}}
\end{center}

\textit{Numerical sample complexity.---} In this appendix, we emphasize that accurate numerical estimates of the Wasserstein-1 distance can, in general, be obtained using a reasonable sample size $R$. Theorem 2 of Ref.~\onlinecite{Fournier2015} implies that for finite $R$, the empirical and exact CDFs, $\hat{P}_{\rm PE}^{(R)}$ and $P_{\rm PE}$, for the projected ensemble are themselves close in Wasserstein-1 distance with high probability: $\mathbbm{P}(W_1(\hat{P}_{\rm PE}^{(R)}, P_{\rm PE}) > \epsilon) \leq e^{-c R\epsilon^2}$ for a constant $c$.
Combined with the triangle inequality
$W_1(\hat{P}_{\rm PE}^{(R)},P_{\rm GHE}) \leq W_1(\hat{P}_{\rm PE}^{(R)},P_{\rm PE}) + W_1(P_{\rm PE},P_{\rm GHE})$ (Ref.~\onlinecite{Villani2009}, Ch.~6), this implies that our estimate will be $\epsilon$-close to the true value $W_1(P_{\rm PE},P_{\rm GHE})$ with probability at least $1-\delta$ provided $R \geq c' \epsilon^{-2}\log (1/\delta)$ for a constant~$c'$.
Given that $W_1(P_{\rm PE},P_{\rm GHE})$ is observed to decay as $1/\textrm{poly}(L_B)$, we can get an arbitrarily small relative error using $R = \text{poly}(L_B)$. For the data in Fig.~\ref{fig:W1_fig}(a), suggesting $W_1(P_{\rm PE},P_{\rm GHE}) \sim L_B^{-1/2}$, taking $R \propto L_B$~suffices.	
\newpage

\begin{onecolumngrid}
    
\newpage
\begin{center}
    {\fontsize{11}{11}\selectfont
        \textbf{Supplemental Material for  ``Matchgate Circuits Deeply Thermalize''\\[5mm]}}
    {\normalsize Mircea Bejan, Benjamin B\'eri, Max McGinley \\[1mm]}
    
\end{center}
\normalsize\

\setcounter{equation}{0}
\setcounter{figure}{0}
\setcounter{table}{0}
\setcounter{page}{1}

\renewcommand{\theequation}{S\arabic{equation}}
\renewcommand{\thefigure}{S\arabic{figure}}
\renewcommand{\thesection}{S\arabic{section}}

\section{Covariance matrix formalism for Gaussian fermionic states \label{app:Gaussian}}

In this appendix, we review the covariance matrix (CM) formalism used for Gaussian fermionic states. We mainly follow Ref.~\onlinecite{bravyi2004gaussian}. 

Consider a system of $L$ fermionic modes with creation and annihilation operators $\hat{c}^\dagger_j$ and $\hat{c}_j$ for $j = 1, \dots, L$ with anticommutation relations $\{ \hat{c}_i, \hat{c}^\dagger_j\} = \delta_{ij}$ and $\{ \hat{c}_i, \hat{c}_j \} = 0$. We denote the fermionic vacuum state by $\ket{0^{\otimes L}}$ in the particle occupation number (i.e. Fock) basis such that $\hat{c}_j\ket{0^{\otimes L}} = 0$ for all $j$. 
It is useful to define the Hermitian Majorana fermion operators $\gamma_{2j-1} \coloneq \hat{c}^\dagger_j + \hat{c}_j$ and $\gamma_{2j} \coloneq\iu(\hat{c}^\dagger_j -\hat{c}_j)$ satisfying $\{ \gamma_i, \gamma_j \} = 2 \delta_{ij}$. 

Any Gaussian fermionic state is fully characterized by its covariance matrix, which contains equal time 2-point correlation functions. For a (possibly mixed) Gaussian state with density matrix $\rho$, the  CM is defined by
\begin{align}
    \Gamma_{jk} \coloneq \mathrm{Tr} \left( \frac{\iu [\gamma_j, \gamma_k]}{2}  \rho \right), \ j,k = 1, \dots, 2L.
\end{align}%
The CM is a real antisymmetric $2L \times 2L$ matrix $\Gamma = \Gamma^* = - \Gamma^T$. For pure states, it satisfies $\Gamma^2 = -\mathbbm{1}$, while for mixed states $\Gamma^2 \ge -\mathbbm{1}$, i.e. the eigenvalues of $\Gamma^2$ are greater than or equal to $-1$.
As an example, the CM of the vacuum state is $\Gamma_0 \coloneq \Gamma ( \ket{0^{\otimes L}} )  = \bigoplus_{i=1}^{L} \begin{psmallmatrix}
    0 & 1 \\ -1 & 0  
\end{psmallmatrix}.$

The unitary evolution of a Gaussian state under any Hamiltonian $H$ quadratic in the operators $\hat{c}^\dagger_j, \hat{c}_j$ preserves the Gaussianity.
Such a Hamiltonian can be written as
\begin{align}
    H = -\frac{\iu}{4} \sum_{j,k=1}^{2L} h_{jk} \gamma_j \gamma_k,
\end{align}%
for an arbitrary real antisymmetric $h = h^* = -h^T$ (up to a constant energy shift in $H$). The effect of evolving under the unitary $U = e^{\iu H}$ on 
a many-body pure Gaussian state $\ket{\Psi}$, the Majorana operators, and the CM is~\cite{Kitaev_unpaired_2001}
\begin{align}
    \ket{\Psi} &\mapsto U \ket{\Psi} = \exp \left( \frac{1}{4} \sum_{j,k=1}^{2L} h_{jk} \gamma_j \gamma_k \right) \ket{\Psi}, \\
    \gamma_j &\mapsto U \gamma_j U^\dagger = \sum_{k=1}^{2L} R_{jk} \gamma_k, \\
    \Gamma &\mapsto R \Gamma R^T,
\end{align}%
where $R = e^h$ is a real orthogonal matrix $R \in \mathrm{SO}(2L)$, i.e. $RR^T = \mathbbm{1}, \det R = 1$.
By choosing the interactions characterized by $h$, 
arbitrary rotations $R \in \mathrm{SO}(2L)$ can be achieved.
We thus note that the local and global gates from Fig.~\ref{fig:setup}(a) are obtained by sampling $R$ from the Haar measure on $\mathrm{SO}(4)$ and $\mathrm{SO}(2L)$, respectively~\cite{Mezzadri_random_matrices}. 

Measuring a Gaussian fermionic state in the single-mode particle occupation basis $\hat{n}_j$ preserves the Gaussianity, where $\hat{n}_j= \hat{c}^\dagger_j \hat{c}_j= \frac{ 1 +\iu\gamma_{2j-1} \gamma_{2j}}{2}$ for $j=1,\dots, L$.
The probability of measuring the occupation $\hat{n}_j$ for a state with CM $\Gamma$ and obtaining outcome $m \in \{ 0,1 \}$ is~\cite{bravyi2004gaussian}
\begin{align}
    p_m = \frac{1}{2 }\sqrt{\det \left( \mathbbm{1} - K^{[j]}_m \Gamma \right)}, \label{eq:pm_single}
\end{align}%
where the $2L \times 2L$ matrix $K^{[j]}_m$ has entries
\begin{align}
    (K^{[j]}_m)_{pq} = (-1)^m (\delta_{p, 2j-1} \delta_{q, 2j} - \delta_{q, 2j-1} \delta_{p, 2j}).\label{eq:K_j_m}
\end{align}The post-measurement Gaussian state has CM
\begin{align}
    \Gamma_m = K^{[j]}_m + D (1-\Gamma K^{[j]}_m)^{-1} \Gamma D, \label{eq:Gamma_m_single}
\end{align}%
where $D_{2j-1, 2j-1} = D_{2j,2j} = 1$ are the only non-zero entries of the $2L \times 2L$ matrix $D$.

It follows that measuring all modes of any subsystem in the Fock basis also yields a Gaussian state.
For simplicity, we focus on the setup yielding the projected ensemble [Fig.~\ref{fig:setup}(a)]. We consider a subsystem $A$ with modes $1, \dots, L_A$ and its complement $B$ with $L_B = L - L_A$ modes.
The probability of measuring all modes in $B$ in the Fock basis with outcomes $\mathbf{m} = (m_1, \dots, m_{L_B})$ is
\begin{align}
    p_\mathbf{m} = \frac{1}{2^{L_B}}\sqrt{\det \left( \mathbbm{1} - K_\mathbf{m} \Gamma \right)}, \label{eq:pm_unblocked}
\end{align}%
and the corresponding post-measurement CM is
\begin{align}
    \Gamma_\mathbf{m} = K_\mathbf{m} + D_\mathbf{m} (1-\Gamma K_\mathbf{m})^{-1} \Gamma D_\mathbf{m}, \label{eq:Gamma_m_unblocked}
\end{align}%
with $K_\mathbf{m} = \mathds{O}_{2L_A} \oplus \kappa_\mathbf{m}$, $\kappa_\mathbf{m} = \bigoplus_{i=1}^{L_B} (-1)^{m_i} \begin{psmallmatrix}
    0 & 1 \\ -1 & 0
\end{psmallmatrix}$, and $D_\mathbf{m} = \mathds{O}_{2L_A} \oplus \mathbbm{1}_{2L_B}$, where $\mathds{O}_d$ is a $d \times d$ matrix of zeroes.
We can simplify Eqs.~(\ref{eq:pm_unblocked}, \ref{eq:Gamma_m_unblocked}) by using the block decomposition of the pre-measurement CM
\begin{align}
    \Gamma = \begin{psmallmatrix}
        \Gamma^A & \Gamma^{AB} \\
        -(\Gamma^{AB})^T & \Gamma^B 
    \end{psmallmatrix}, \label{eq:Gamma_block}
\end{align} where $\Gamma^A$ is the $2L_A \times 2L_A$ CM of the state with the reduced density matrix $\rho^A = \Tr_B (\rho)$, and $\Gamma^{AB}$ is a $2L_A \times 2L_B$ matrix capturing the correlations between subsystems $A$ and $B$.
This block decomposition also holds for the post-measurement CM with $\Gamma^{AB}_\mathbf{m} = \mathds{O}$, $\Gamma^B_\mathbf{m} = \kappa_\mathbf{m}$,
\begin{align}
    \Gamma^A_\mathbf{m} = \Gamma^A + \Gamma^{AB} \kappa_\mathbf{m} (\mathbbm{1}-\Gamma^B \kappa_\mathbf{m})^{-1} (\Gamma^{AB})^T, 
    \label{eq:Gamma_m_blocked}
\end{align}%
and we can rewrite Eq.~(\ref{eq:pm_unblocked}) as 
\begin{align}
    p_\mathbf{m} = \frac{1}{2^{L_B}}\sqrt{\det \left( \mathbbm{1} - \kappa_\mathbf{m} \Gamma^B \right)}.\label{eq:pm_blocked}
\end{align}

\section{Sampling from the projected ensemble \label{app:sampling}}
In this appendix, we present the sampling procedure for constructing an empirical projected ensemble. 
While Eqs.~(\ref{eq:Gamma_m_blocked}, \ref{eq:pm_blocked}) are useful for analytical calculations, iterative procedures based on single-mode measurements [such as Eqs.~(\ref{eq:pm_single}, \ref{eq:Gamma_m_single})] are better suited to numerically sample $\Gamma^A_\mathbf{m}$ with probability $p_\mathbf{m}$. One such procedure goes as follows.
We denote the pre-measurement CM by $\Gamma^{(0)}\coloneq \Gamma$. First, we sample~$\Gamma^{(1)}\coloneq \Gamma({m_1})$ according to $p(m_1) = \sqrt{\det \left( 1 - K^{(1)} \Gamma^{(0)} \right)}/2$, where $K^{(j)} = K^{[L_A + j]}_{m_j}$, cf. Eq.~(\ref{eq:K_j_m}).
Then, we sample $\Gamma^{(2)} \coloneq \Gamma({m_2 \vert m_1})$ according to the conditional probability $p(m_2 \vert m_1) = \sqrt{\det \left( 1 - K^{(2)} \Gamma^{(1)} \right)}/2$.
Further, we continue sampling $\Gamma^{(j)} \coloneq \Gamma(m_j \vert {m_1, \dots ,m_{j-1}})$~with 
\begin{align}
    p(m_j \vert m_1, \dots, m_{j-1}) = \frac{1}{2}\sqrt{\det \left( 1 - K^{(j)} \Gamma^{(j-1)} \right)}
\end{align}until we sample $\Gamma^{(L_B)}$ as given by the conditional $p(m_{L_B} \vert m_1, \dots , m_{L_B -1})$.
Hence, by chaining the conditional probabilities, we have sampled $\Gamma_\mathbf{m}$ with
\begin{align}
    p_\mathbf{m} = p(m_1) p(m_2 \vert m_1) \cdots 
                   p(m_{L_B} \vert m_1, \dots , m_{L_B - 1}).
\end{align}%
Finally, we obtain $\Gamma^A_\mathbf{m}$ by taking the relevant block of $\Gamma_\mathbf{m}$.

We note that using a single-mode version of Eqs.~(\ref{eq:Gamma_m_blocked}, \ref{eq:pm_blocked}) is  numerically more efficient than the approach based on Eqs.~(\ref{eq:pm_single}, \ref{eq:Gamma_m_single}) presented above. This is because the former only keeps track of the unmeasured system up to a certain stage in the sampling procedure and involves smaller matrices for computing the conditional probabilities. We use these more efficient equations in our simulations but omit their presentation for brevity.

\section{Singular values of subsystem covariance matrices}
In this appendix, we review how the singular values of a subsystem CM characterize the ensemble of CM through their joint PDF. We derive this joint PDF using a mapping between CM and scattering matrices. This mapping allows us to utilize results from the theory of quantum transport and, thus, provides a simpler derivation than that of Ref.~\onlinecite{bianchi_pageFGS} where recent results in random matrix theory were used. Then, using this general formula for the joint PDF, we derive the CDF of a CM element. We also review how these singular values enter the calculation of the entanglement entropy, and for completeness, mention the mean and variance of the EE in the GHE.

\subsection{Definition of the singular values}
Here, we focus on the covariance matrices of pure Gaussian fermionic states on $L$ modes. We consider two subsystems $A$ and $B$ consisting of the first $L_A$ and last $L_B = L-L_A$ modes, respectively, with $L_A \leq L_B$ w.l.o.g. Thus, $\Gamma$ has the block decomposition of Eq.~(\ref{eq:Gamma_block}). 

We are interested in the singular values of the subsystem CM $\Gamma^A$. Since $\Gamma$ is real antisymmetric, we can perform a singular value decomposition of $\Gamma^{AB} = O \Lambda Q^T$, where $\Lambda$ is a $2L_A \times 2L_B$ matrix containing the two-fold degenerate singular values $\lambda_i \in [0, 1]$ of $\Gamma^{AB}$ for $i=1, \dots, L_A$ 
\begin{align}
    \Lambda = \begin{psmallmatrix}
        \lambda_1 & & & & & &\vert & \\
         & \lambda_1 & & & & &\vert & \\
         & & \lambda_2 & & & &\vert & \mathds{O} \\
        & & & \dots & & &\vert & \\ 
        & & & & & \lambda_{L_A} &\vert & \\ 
    \end{psmallmatrix},
\end{align}%
and $O, Q$ are real orthogonal matrices~\cite{SchuchBauer_prb19}. Since $\Gamma$ is pure $\Gamma^2 = -\mathbbm{1}$, we can use the same $O, Q$ to rewrite $\Gamma$ as
\begin{align}
    \Gamma = \begin{psmallmatrix}
        O & \\ & Q \\
    \end{psmallmatrix}
    \begin{psmallmatrix}
        K_A \sqrt{\mathbbm{1}-\Lambda \Lambda^T} & \Lambda \\ -\Lambda^T & K_B \sqrt{\mathbbm{1} - \Lambda^T \Lambda}
    \end{psmallmatrix}
    \begin{psmallmatrix}
        O^T & \\ & Q^T \\
    \end{psmallmatrix}, \label{eq:Gamma_svd}
\end{align}where $K_X = \bigoplus_{i=1}^{L_X} \begin{psmallmatrix}
    0 & 1 \\ -1 & 0
\end{psmallmatrix}$. %
Hence, the two-fold degenerate singular values $s_i \in [0,1]$ of $\Gamma^A$ can be related to those of $\Gamma^{AB}$ by $s_i = \sqrt{1-\lambda_i^2}$ for $i =1,\dots,L_A$. 
The decomposition in Eq.~(\ref{eq:Gamma_svd}) is also useful for computing entanglement measures~\cite{SchuchBauer_prb19}, see also App.~\ref{app:EE}.

More precisely, we are concerned with the joint probability distribution of these singular values $p(\{ s_i \})$. This joint PDF\footnote{In fact, there is one joint PDF $p(\{ s_i \})$ for each $L_A = 1,\dots, L-1$.} characterizes the GHE since the GHE is equivalent to that of random covariance matrices $\Gamma \in \frac{\mathrm{O}(2L)}{U(L)}$. We shall next provide a derivation of this joint PDF.

\subsection{Derivation of the joint PDF via mapping CM to scattering matrices \label{app:Scattering}}

One can regard the CM $\Gamma$ of a pure state in the GHE as a scattering matrix $\mathcal{S} \coloneq \Gamma$ with $\mathcal{S} \mathcal{S}^\dagger = \mathbbm{1}$ because $\Gamma$ is real orthogonal $\Gamma \Gamma^T = \mathbbm{1}$. 
We note that Ref.~\onlinecite{Jian2022} also mapped the CM of a Gaussian fermionic state to a scattering problem: they map the CM of the Choi state $\Gamma_U$ corresponding to a Gaussian unitary $U$ to a scattering matrix; here, we consider instead the CM of the states on which such a unitary $U$ would act on.

For a bipartition into contiguous\footnote{One can also construct a scattering problem with a local scattering center at the boundary between subsystems $\mathcal{A,B}$ for arbitrary, i.e. not necessarily contiguous, subsystems $A,B$. However, this locality in the scattering problem does not imply that the boundary between $A$ and $B$ is local and connected.} subsystems $A$ and $B$, the associated scattering problem [Fig.~\ref{fig:scatt}] consists of $N_\mathcal{A} = 2L_A$ ($N_\mathcal{B} =2L_B$) modes\footnote{These are also referred to as scattering channels~\cite{beenakker_RMP}.} propagating
within subsystem $\mathcal{A}$ ($\mathcal{B}$) and scattering at the interface between $\mathcal{A}$ and $\mathcal{B}$. The basis we choose for these modes is $\ket{\varphi^{\mathcal{A},+}_j} = \ket{e_j} + \ket{h_j}$ and $\ket{\varphi^{\mathcal{A},-}_j} = \iu (\ket{e_j} - \ket{h_j})$ with $j=1,\dots, N_\mathcal{A}$, where $\ket{e_j}$ ($\ket{h_j}$) is the electron (hole) state of the $j$th channel~\cite{dahlhaus_prb_2010,Jian2022}; a similar definition holds for subsystem~$\mathcal{B}$. Note that the creation and annihilation operators for each state $\ket{\varphi^{\mathcal{A(B)}, \pm}_j}$ are Hermitian, thus, Majorana operators.

The scattering matrix $\mathcal{S}$ relates these channels by
\begin{align}
    \begin{psmallmatrix}
        \varphi^{\mathcal{A},-} \\
        \varphi^{\mathcal{B},+}
    \end{psmallmatrix}
    = \mathcal{S}
    \begin{psmallmatrix}
        \varphi^{\mathcal{A},+} \\
        \varphi^{\mathcal{B},-}
    \end{psmallmatrix}, \label{eq:S_phis}
\end{align}%
where the amplitudes on the RHS (LHS) of Eq.~(\ref{eq:S_phis}) correspond to incoming (outgoing) modes to (from) the scattering center.
It is instructive to rewrite $\mathcal{S}$ using the block decomposition from Eq.~(\ref{eq:Gamma_block})
\begin{align}
    \mathcal{S} = \begin{pmatrix}
        \mathcal{R} & \mathcal{T}' \\
        \mathcal{T} & \mathcal{R}' 
    \end{pmatrix}, \label{eq:S_block}
\end{align}%
where $\mathcal{R} = \Gamma^A$, $\mathcal{T}' = \Gamma^{AB}$, etc. are $N_\mathcal{A} \times N_\mathcal{A}$, $N_\mathcal{A} \times N_\mathcal{B}$, etc. matrices, respectively. Hence, we can identify $\mathcal{R}$ as a reflection matrix for the modes in $\mathcal{A}$ and $\mathcal{T}'$ as a transmission matrix of modes from $\mathcal{B}$ to $\mathcal{A}$. The transmission eigenvalues $T_i \in [0,1]$ of $\mathcal{T}' \mathcal{T}'^{\dagger}$ and reflection eigenvalues $R_i = 1-T_i \in [0,1]$ of $\mathcal{R} \mathcal{R}^\dagger$ for $i = 1, \dots , \min(N_\mathcal{A}, N_\mathcal{B})$ are central to the theory of quantum transport, e.g., for computing the conductance~\cite{beenakker_RMP}.

\begin{figure}[t]
    \centering
        \includegraphics[width=8.6cm]{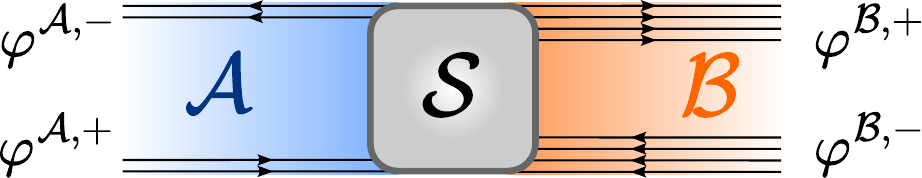}%
    \caption{%
             Unitary scattering problem at the interface between subsystems $\mathcal{A}$ and $\mathcal{B}$. The scattering matrix $\mathcal{S}$ relates the Majorana modes $\varphi^{\mathcal{A(B)}, \pm}$ through Eq.~(\ref{eq:S_phis}).
            }
    \label{fig:scatt}
\end{figure}

The probability distribution of transmission, and thus reflection, eigenvalues is known for various symmetry classes of scattering matrices~\cite{dahlhaus_prb_2010, beenakker_RMP} and, from it, we can infer $p(\{ s_i \})$.
In our case, the scattering matrix $\mathcal{S}$ inherits symmetries from the properties of $\Gamma$: it has particle-hole symmetry $\mathcal{S}=\mathcal{S}^*$, time-reversal symmetry $\mathcal{S}=-\mathcal{S}^T$ and hence chiral symmetry $\mathcal{S}^2 = - \mathbbm{1}$, cf. Ref.~\onlinecite{Fulga_prb_2011}. These symmetries place $\mathcal{S}$ in the Altland-Zirnbauer symmetry class DIII~\cite{AltlandZirnbauer, Fulga_prb_2011}. 
For the symmetry class DIII, $R_i$ are two-fold degenerate, and the joint PDF of the (possibly) non-degenerate $L_A$ (assuming $L_A \leq L_B$) of them is~\cite{dahlhaus_prb_2010}
\begin{align}
    p(\{ R_i \}) \propto \prod_{i=1}^{L_A} \frac{(1-R_i)^{\vert L_A - L_B\vert}}{\sqrt{R_i}} \prod_{1 \leq j < k \leq L_A} \vert R_j - R_k \vert^2. 
\end{align}Note that since $\mathcal{R} = \Gamma^A$, we have $s_i = \sqrt{R_i}$ for all $i$. Thus, by a change of variables, we obtain the joint PDF of the (possibly) non-degenerate singular values
\begin{align}
    p(\{ s_i\}) = \mathcal{N} \prod_{i=1}^{L_A} (1-s^2_i)^{\vert L_A - L_B\vert} \prod_{1 \leq j < k \leq L_A} \vert s^2_j - s^2_k \vert^2, \label{eq:p_svals}
\end{align}with normalization $\mathcal{N}$. This concludes our derivation.

\subsection{PDF and CDF of a CM element \label{app:CDFx}}
We next obtain the probability density of any off-diagonal CM element by considering on a special case of Eq.~(\ref{eq:p_svals}).  Without loss of generality, we can focus on the CM element $x \coloneq \Gamma_{12} \in [-1,1]$ due to the Haar invariance.
By fixing $L_A=1$ and $L_B = L-1$, we have $\Gamma^A = \begin{psmallmatrix}
    0 & x \\ -x & 0 
\end{psmallmatrix}$, and thus $x = \pm s$, where $s$ is the only (two-fold degenerate) singular value of $\Gamma^A$. The sign in $x = \pm s$ is determined by the parity sector of $\Gamma^A$ and both parity sectors are equally probable by Haar invariance. Thus, a direct application of Eq.~(\ref{eq:p_svals}) yields the desired PDF
\begin{align}
    p_\mathrm{GHE}(x) = \frac{\mathcal{N}}{2} (1-x^2)^{L_B-1}, \quad x \in [-1, 1].
\end{align}We can further obtain the cumulative distribution of $x$
\begin{align}
    P_\mathrm{GHE}(x) &= \int^x_{-1} \dif u \, p_\mathrm{GHE}(u) \\
         &= \frac{1}{2} + \frac{x \Gamma(L_B + \frac{1}{2})}{\sqrt{\pi} \Gamma(L_B)}  \ _2F_1 \left( \frac{1}{2}, -L_B + 1, \frac{3}{2}; x^2 \right),
\end{align}where $\Gamma(z)$ and $_2F_1 (a, b, c; z)$ are the gamma and hypergeometric functions, respectively.

\subsection{Entanglement entropy \label{app:EE}}
The singular values $\{s_i\}$ of $\Gamma^A$ are also useful for calculating entanglement measures. Using a fermionic Gaussian analogue of the Schmidt decomposition~\cite{BOTERO200439,SchuchBauer_prb19}, the von Neumann EE between subsystems $A$ and $B$ is given by 
\begin{align}
    S(\rho^A) = \sum_{i=1}^{L_A} &-\left(\frac{1-s_i}{2} \right)\log \left( \frac{1-s_i}{2} \right)  -\left(\frac{1+s_i}{2}\right) \log \left( \frac{1+s_i}{2} \right).
    \label{eq:vonNeumannSV}
\end{align}

The exact expression of $p(\{ s_i \})$ enabled the calculation of the ensemble mean and variance EE~\cite{bianchi_pageFGS}. For convenience, we list these ensemble averages below since we used them in our numerics. The mean EE is defined as
\begin{align}
    \langle S(f) \rangle_\mathrm{GHE}\coloneq \underset{\Gamma \sim \Upsilon_{\rm pure}}{\mathbbm{E}} S(\rho^A),
\end{align}where $f \coloneq \frac{L_A}{L}$; we denote as $\Upsilon$ the space of covariance matrices of (possibly mixed) Gaussian fermionic states on $L$ modes, and $\Upsilon_\mathrm{pure} = \mathrm{SO}(2L) / \mathrm{U}(L) \subset \Upsilon$ the subset corresponding to pure states. Similarly, the ensemble EE variance is 
\begin{align}
    \left[ \Delta S(f)_\mathrm{GHE}\right]^2 \coloneq \underset{\Gamma \sim \Upsilon_{\rm pure}}{\mathbbm{E}} \left[ S(\rho^A)  - \langle S(f) \rangle_\mathrm{GHE}\right]^2.
\end{align}%
The exact expression for the mean EE reads~\cite{bianchi_pageFGS}
\begin{align}
    \langle S(f) \rangle_\mathrm{GHE} = & \left( L - \frac{1}{2} \right) \psi(2L) + \left(  \frac{1}{4} -L_A  \right) \psi(L) + \left( \frac{1}{2} +L_A-L\right) \psi(2L-2L_A) 
    + \frac{1}{4}\psi(L - L_A) -L_A,
\end{align}where $\psi(z)$ is the digamma function, while the variance of the EE in the thermodynamic limit is
\begin{align}
    \lim_{L \to \infty} \left[ \Delta S(f)_\mathrm{GHE}\right]^2 = \frac{f+f^2+\log(1-f)}{2}.
\end{align}

\section{Proof of Theorems \ref{thm:Moments} and \ref{thm:MomentsManyBody} \label{app:Thm1proof}}
	
	In this appendix, we prove Theorems \ref{thm:Moments} and \ref{thm:MomentsManyBody}, which state that for sufficiently large $L_B$, the $k$th moments of the covariance matrices \eqref{eq:MomentMatrixElement} and many-body states \eqref{eq:MomentManyBody} in projected ensemble are with high probability close to those of the Gaussian Haar ensemble. The results in question will follow relatively quickly from the following lemma.
	\begin{lemma} \label{lemma:FConcentration}
		Let $f :\Upsilon^A \rightarrow \mathbbm{R} $ 
        be a real-valued, once-differentiable function of covariance matrices of (possibly mixed) Gaussian fermionic states on $A$, which is $c$-Lipschitz and bounded in an interval $[a,b]$,
		\begin{align}
			f(\Gamma^A) \in [a,b] \;\;\text{ and }\;\;\sum_{ij}\left|\frac{\partial f(\Gamma^A)}{\partial [\Gamma^A]_{ij}}\right|^2 \leq c^2, \hspace*{12pt}\forall\, \Gamma^A \in \Upsilon^A.
			\label{eq:FSmoothThm1}
		\end{align}
		Then, there exists a constant
  \begin{align}
      \kappa_f \geq \frac{1}{16} \cdot \frac{1}{[c + (b-a)/4]^2}
  \end{align}
  such that
		\begin{align}
			\mathbbm{P}_{O \sim \textup{Haar}}\left(\left|\sum_{\mathbf{m}} p_{\mathbf{m}} f(\Gamma^A_{\mathbf{m}}) - \langle f (\Gamma^A) \rangle_{\Gamma^A \sim \textup{Haar}(\Upsilon^A)} \right| > \epsilon \right) %
			\leq 2\exp\left(-\kappa_f L_B \epsilon^2\right).
			\label{eq:FConcentration}
		\end{align}
	\end{lemma}
\noindent The proof of Lemma~\ref{lemma:FConcentration} is given in the next section, i.e., App.~\ref{app:lemma1_proof}. From this result, we now prove Theorems \ref{thm:Moments} and \ref{thm:MomentsManyBody}.

 \subsection{Proof of Theorem~\ref{thm:Moments}}
 For each pair of multi-indices $\mathbf{i}$, $\mathbf{j}$, the corresponding  moment $\mathcal{C}^{(k)}_{\rm PE, \mathbf{i}\mathbf{j}}$ [Eq.~\eqref{eq:MomentMatrixElement}] is given by the average of the function
	\begin{align}
		\mathcal{B}^{(k)}_{\mathbf{i}\mathbf{j}}(\Gamma^A) = \prod_{a=1}^k [\Gamma^A]_{i_a j_a}, 
	\end{align}
	over covariance matrices sampled from the projected ensemble $\Gamma^A \sim \{(\Gamma^A_\mathbf{m}, p_\mathbf{m})\}$. The function $\mathcal{B}^{(k)}_{\mathbf{i} \mathbf{j}}(\Gamma^A)$  satisfies Eq.~\eqref{eq:FSmoothThm1} with $c \leq k$ and $[a,b] = [-1,1]$. Combining Eq.~\eqref{eq:FConcentration} with $\kappa_f \geq \frac{1}{16(k + 1/2)^{2}}\geq \frac{1}{64k^2}$, we arrive at Eq.~\eqref{eq:MomentConcentrate}.

 \subsection{Proof of Theorem~\ref{thm:MomentsManyBody}}
 To prove Theorem \ref{thm:MomentsManyBody}, we require the following dual characterization of the trace distance between the $k$th moments of the many-body states
\begin{align}
         \Delta^{(k)} = \sup_{O^{(k)} : \|O^{(k)}\|_\infty \leq 1} \Tr\left[O^{(k)}\Big(\mathcal{E}^{(k)}_{\textrm{PE}} - \mathcal{E}^{(k)}_{\textrm{GHE}} \Big)\right] \leq& \sup_{c_S : \sum_S |c_S|^2 \leq 1} \sum_{S} c_S \Tr\Big[\pi^S\big(\mathcal{E}^{(k)}_{\textup{PE}} - \mathcal{E}^{(k)}_{\textup{GHE}}\big)\Big] \nonumber\\ \eqqcolon& \sup_{c_S : \sum_S |c_S|^2 \leq 1} \sum_{S} c_S \Delta_S.
         \label{eq:DeltaKDual}
     \end{align}
     Here, $\{\pi^S\}$ is the set of all possible products of Majorana operators on $k$ copies of a $L_A$-mode fermionic system, namely operators of the form
     \begin{align}
         \pi^S =\bigotimes_{a=1}^k \left( \prod_{j \in S_a} \gamma_j \right)
         \label{eq:PiSDef}
     \end{align}
     where the label $S$ is given by a $k$-tuple $S = (S_1, \ldots, S_k)$, with each element being a subset $S_a \subseteq [2L_A]$. For the inequality in Eq.~\eqref{eq:DeltaKDual}, we used
         $\Vert O^{(k)} \Vert^2_\infty \geq D^{-1} \Vert O^{(k)} \Vert^2_2 = D^{-1} \sum_{S,R} c^*_S c_R \Tr [ \pi^\dagger_S \pi_R] = \sum_S |c_S|^2$, with the replicated Hilbert space dimension $D$. 
     Since the moment operators $\mathcal{E}^{(k)}_{\rm PE}$ commute with the fermion parity operator on each copy, we can take each $S_a$ to contain an even number of fermions in the above. Accordingly, the $k$-copy expectation value can be evaluated using Wick's theorem \cite{Surace2022}
     \begin{align}
         \mathcal{B}^S(\Gamma^A) \coloneqq \Tr\left[\iu^{-|S|/2}\pi^S \rho(\Gamma^A)^{\otimes k}\right] = \prod_{a=1}^k \Pf(\Gamma^A_{S_a}),
         \label{eq:ProductCorrelator},
     \end{align}
     where $\rho(\Gamma^A)$ is a Gaussian state with covariance matrix $\Gamma^A$, $|S| = \sum_a |S_a|$, and $\Gamma^A_{S_a}$ is the submatrix of $\Gamma^A$ containing the rows and columns corresponding to the Majorana operators contained in $S_a$. By definition, the function $\mathcal{B}^S(\Gamma^A)$ is bounded in $[-1,1]$, and we can also evaluate its Lipschitz constant
     \begin{align}
         c^2 = \sup_{\Gamma^A \in \Upsilon^A} \sum_{ij} \left| \frac{\partial \mathcal{B}^S}{\partial [\Gamma^A]_{ij}} \right|^2 = \sup_{\Gamma^A \in \Upsilon^A} \|\nabla \mathcal{B}^S\|_2^2,
     \end{align}
     where $\nabla \mathcal{B}^S$ is the $2L_A \times 2L_A$ matrix of derivatives, with elements $[\nabla \mathcal{B}^S]_{ij} = \partial \mathcal{B}^S/\partial [\Gamma^A_{ji}]$. Using Jacobi's formula for the Pfaffian, $\nabla_X \Pf(X) = \frac{1}{2}\Pf(X) X^{-1}$, we have
     \begin{align}
         \nabla \mathcal{B}^S = \frac{1}{2}\sum_{\substack{a=1\\ i \in S_a \wedge j \in S_a }}^k \Big(\Pf(\Gamma^A_{S_a})\cdot(\Gamma^A_{S_a})^{-1} \oplus \mathbbm{O}_{[2L_A] \backslash S_a} \Big)\prod_{\substack{b=1\\ b \neq a}}^k \Pf(\Gamma^A_{S_b}).
     \end{align}
     Now, $\Gamma^A_{S_a}$ is a real antisymmetric, so one can always find a transformation $R \in \textrm{SO}(2L_A)$ that brings it into the standard form
 \begin{align}
     \Gamma^A_{S_a} = R \bigoplus_{r=1}^{|S_a|/2} \begin{pmatrix}
         0 & \lambda_r  \\
         -\lambda_r & 0
     \end{pmatrix}R^T.
 \end{align}
 By the identity $\Pf(R A R^T) = \det(R) \Pf(A)$, we can  write the Pfaffian as $\Pf(\Gamma^A_{S_a}) = \prod_{r = 1}^{|S_a|/2} \lambda_r$. Moreover, since $\|\Gamma^A\|_\infty \leq 1$ and $\Gamma^A_{S_a}$ is a submatrix of $\Gamma^A$, we have $|\lambda_r| \leq 1$ for all $r$. Now, assuming that $\Gamma^A_{S_a}$ is non-singular, we can write
 \begin{align}
     \Pf(\Gamma^A_{S_a})[(\Gamma^A_{S_a})^{-1}]_{ij} &= \left[ R \bigoplus_{r=1}^{|S_a|/2} \begin{pmatrix}
         0 &  \tilde{\lambda}_{r}  \\
         -\tilde{\lambda}_r & 0
     \end{pmatrix}R^T\right]_{ij}, & \text{where }\tilde{\lambda}_r = \prod_{r' \neq r} \lambda_{r'}.
     \label{eq:PfaffianInverse}
 \end{align}
 If some of the eigenvalues $\lambda_r$ are zero, we can shift them by a small amount to ensure that all $\Gamma^A_{S_a}$ is non-singular, and then take the limit as some $\lambda_r$ tend to zero,  which is well-defined if we use the right hand side of Eq.~\eqref{eq:PfaffianInverse}. 
 Using H{\"o}lder's inequality and $\| R \|_\infty=1$, this gives us $\|\Pf(\Gamma^A_{S_a}) \cdot (\Gamma^A_{S_a})^{-1}\|_2 \leq \sqrt{2L_A}$. Thus, by the triangle inequality, we have $c = \sup_{\Gamma^A \in \Upsilon^A} \|\nabla \mathcal{B}^S\|_2 \leq k\sqrt{L_A/2}$. 
 Hence, from Lemma \ref{lemma:FConcentration}, we have that for any $S$,
 \begin{align}
     \mathbbm{P}_{O \sim \textrm{Haar}}\left(
     \Delta_S
     > \epsilon \right) \leq 2 \exp\left(-\frac{L_B\epsilon^2}{16L_A k^2}\right).
 \end{align}
 Combining the above with Eq.~\eqref{eq:DeltaKDual}, we have that $\Delta^{(k)}$ can only exceed $\epsilon$ if there is some set of numbers $c_S$ satisfying $\sum_S |c_S|^2 \leq 1$ for which $\sum_S c_S \Delta_S > \epsilon$. Because there are at most $2^{2(L_A-1)k}$ terms in this sum (recall that we only include operators that are fermion parity even on each copy), this means that for at least one such $S$, we will have $\Delta_S > 2^{-L_Ak}\epsilon$. Applying a union bound, we get
 \begin{align}
     \mathbbm{P}_{O \sim \textrm{Haar}}\left( \Delta^{(k)} > \epsilon \right) \leq \mathbbm{P}_{O \sim \textrm{Haar}}\left( \bigcup_S\, \Delta_S > 2^{-(L_A-1)k}\epsilon \right) \leq 2^{2(L_A-1)k + 1}\exp\left(-\frac{L_B\epsilon^2}{16 L_Ak^2 2^{(L_A-1)k}} \right).
 \end{align}
 The right hand side of the above is less than $\delta$ for
 \begin{align}
     L_B \geq \frac{16L_Ak^2 2^{(L_A-1)k}}{\epsilon^2}\Big( \log(1/\delta) + \log 2 \times (2(L_A-1)k +1)\Big)
 \end{align}
 Assuming that $\delta \leq 1/2$, the above is satisfied if $L_B \geq 32L_A^2k^3 2^{(L_A-1)k}\log(1/\delta)/\epsilon^2$. In the regime $kL_A \geq 20$, we have $32L_A^2k^3 2^{(L_A-1)k} \leq e^{2L_Ak}$. Hence, the concentration inequality holds if we have
 \begin{align}
     2kL_A \leq \log(L_B) - 2\log(1/\epsilon)- \log \log(1/\delta),
 \end{align}
 and the right hand side of the above exceeds 20. If we impose that $\frac{1}{2}\log(L_B) \geq 20 +2\log (1/ \epsilon) +\log \log(1/\delta)$, then the above becomes $kL_A \leq \frac{1}{4}\log (L_B)$ and this gives us the condition quoted in Theorem \ref{thm:MomentsManyBody}. \hfill $\square$

\subsection{Proof of Lemma \ref{lemma:FConcentration} \label{app:lemma1_proof}}
 
	Let us use the shorthand $\nu_{\textrm{PE}}(f) = \sum_{\mathbf{m}} p_{\mathbf{m}} f(\Gamma^A_{\mathbf{m}})$ and similarly we write $\nu_{\textrm{GHE}}(f) = \int_{\Upsilon^A_{\rm pure}} \dif \nu_{\rm GHE}(\Gamma^A) f(\Gamma^A)$ for the corresponding Gaussian-Haar average. 
Because $f$ can be shifted by a constant without affecting the difference between these averages, we will assume that $f$ is bounded in a symmetric interval
$f \in [-b,b]$---the more general case $f \in [a,b]$ then follows from setting $b \rightarrow (b-a)/2$.
 Recalling that the projected ensemble $\nu_{\rm PE}$ is a function of the orthogonal transformation $O$ used to prepare the pre-measurement state, we claim that $\mathbbm{E}_{O \sim \text{Haar}}\nu_{\textrm{PE}}(f) = \nu_{\textrm{GHE}}(f)$. To see this, note that the Haar distribution over Gaussian transformations $O$ on the whole system [i.e., over $\mathrm{SO}(2L)$] is invariant under arbitrary rotations $O^A$ that only act on $A$, and these operations in turn commute with the measurements on $B$. Thus, we have
	\begin{align}
		\mathbbm{E}_{O \sim \text{Haar}}\nu_{\textrm{PE}}(f) &= \mathbbm{E}_O \sum_{\mathbf{m}} p_{\mathbf{m}} \mathbbm{E}_{O^A} f\Big(O^A \Gamma^A_{\mathbf{m}} (O^A)^T\Big) \nonumber\\
		&= \mathbbm{E}_{O^A} f\Big(O^A \Gamma_0 (O^A)^T\Big) = \nu_{\textrm{GHE}}(f),
		\label{eq:PEHaarAvg}
	\end{align}
	as claimed. In going to the second line, we use the fact that for any pure state covariance matrix $\Gamma_0$, the ensemble of states $O^A \Gamma_0 (O^A)^T$ for Haar-distributed $O^A$ is independent of $\Gamma_0$. Having determined the mean of $\nu_{\rm PE}(f)$ over $O$, we can now use the following concentration of measure result.
	\begin{lemma}[Ref.~\onlinecite{Anderson2009}, Corollary 4.4.28] \label{lemma:Levy}
		Let $g : \mathrm{SO}(n) \rightarrow \mathbbm{R}$ be a function with Lipschitz constant $\eta > 0$, that is for any two $O_1, O_2 \in \mathrm{SO}(n)$ we have
		\begin{align}
			|g(O_1) - g(O_2)| \leq \eta \|O_1 - O_2\|_2
			\label{eq:Lipschitz}
		\end{align}
		where $\|X\|_2 = \sqrt{\Tr[X^\dagger X]}$ is the Frobenius norm over the space of $n\times n$ matrices. If $O$ is distributed according to the Haar measure over $\mathrm{SO}(n)$, then the probability that $g(O)$ deviates from its mean $\langle g \rangle_{\textup{Haar}} = \mathbbm{E}_{O \sim \textup{Haar}} g(O)$ by more than $\epsilon$ is upper bounded as
		\begin{align}
			\mathbbm{P}_{O \sim \textup{Haar}}\left(|g(O) - \langle g \rangle_{\textup{Haar}}| > \epsilon \right) \leq 2 \exp\left(-\frac{(n-2)\epsilon^2}{8\eta^2}\right).
		\end{align}
	\end{lemma}
	In Ref.~\onlinecite{Cotler2023}, a similar kind of concentration bound was used to show that $N$-qubit states generated by Haar-random unitaries exhibit deep thermalization. There, the relevant space of unitaries was $\mathrm{U}(2^N)$, whereas here due to our restriction to Gaussian fermionic states, we have a much smaller group of possible operations $\mathrm{SO}(2L)$. Mirroring the approach of Ref.~\onlinecite{Cotler2023}, our goal now is to show that $\nu_{\textrm{PE}}(f)$, when viewed as a function of $O$, satisfies Eq.~\eqref{eq:Lipschitz} for some appropriate constant $\eta$.
	
	Let $\ket{\Psi(O)}$ be the many-body quantum state (which is a vector in the Fock space) that results from applying the Gaussian evolution $O$ to the initial state $\ket{0^{\otimes L}}$. The function of interest $g(O)$ is the average of $f$ in the projected ensemble for this state,
	\begin{align}
		g(O) = g(\Psi(O)) \coloneqq \nu_{\rm PE}(f)_{\ket{\Psi(O)}} = \sum_{\mathbf{m}} p_{\mathbf{m}}\, f\left( \frac{\tilde{\Gamma}_{\mathbf{m}}^A}{p_{\mathbf{m}}} \right), \label{eq:GODef}
	\end{align}
	where the probabilities are given by
	\begin{align}
		p_{\mathbf{m}} =  \big\langle \Psi(O)\big| \mathbbm{1}_A \otimes P_\mathbf{m}  \big| \Psi(O)\big\rangle
  \label{eq:PostMeasPsi1}
	\end{align}
	and we define the unnormalized covariance matrices of the unnormalized post-measurement states are
	\begin{align}
		[\tilde{\Gamma}^A_{\mathbf{m}}]_{jk} =  \big\langle \Psi(O)\big| \pi_{jk} \otimes P_\mathbf{m}  \big| \Psi(O)\big\rangle,
  \label{eq:PostMeasPsi2}
	\end{align}
	with $\pi_{jk} \coloneqq \frac{\iu}{2}[\gamma_j, \gamma_k]$. In the above, we write the projection operator $P_\mathbf{m} = \ket{\mathbf{m}}\bra{\mathbf{m}}_B$. 

 Because $g(O)$ is differentiable, and the manifold $\mathrm{SO}(2L)$ to which $O$ belongs has positive curvature, its Lipschitz constant $\eta$ [Eq.~\eqref{eq:Lipschitz}] can be determined by considering elements $O_{1,2}$ that are infinitesimally close to one another. We therefore take $O_1 = O$ and $O_2 = e^{\dif M}O = (\mathbbm{1}+\dif M)O$, where $\dif M$ is an infinitesimal element of the Lie algebra $\mathfrak{so}(2L)$, which consists of real antisymmetric matrices. The covariance matrices corresponding to the two states are then given by
 \begin{align}
     \Gamma_1 &= \Gamma \coloneqq O \Gamma_0 O^T & \Gamma_2 &= O \Gamma O^T = \Gamma + [\dif M, \Gamma].
     \label{eq:GammaInfShift}
 \end{align}
 Also, by the invariance of the Frobenius norm under left-multiplication by an orthogonal matrix, we have
	\begin{align}
		\|O_1 - O_2\|_2 = \|\dif M\|_2.
		\label{eq:DifODA}
	\end{align}
	Now, we wish to evaluate $\dif g = g(e^{\dif M}O) - g(O)$. To do this, we use the fact any fermion linear optical transformation, whose single-mode description is equal to $e^{W}$ for a real antisymmetric $W$, can be described by a unitary in Fock space $U(e^W) = e^{\iu \theta}\exp(\frac{1}{4}\sum_{jk} \gamma_j [W]_{jk} \gamma_k)$ \cite{Kitaev_unpaired_2001}, where $\theta$ is an overall phase that does not affect $g(\Psi)$. Thus,
	\begin{align}
		\ket{\Psi(e^{\dif M}O)} = \exp\left(\frac{1}{4}\sum_{jk} \gamma_j [\dif M]_{jk} \gamma_k + \iu \dif \theta\right)\ket{\Psi(O)}
	\end{align}
	where $\dif \theta$ specifies an infinitesimal phase change. Because $\dif M$ can be shifted by any matrix proportional to $\Gamma$ without affecting the covariance matrix [see Eq.~\eqref{eq:GammaInfShift}], we can absorb the choice of $\dif \theta$ into $\dif M$, in such a way that
    \begin{align}
        \Tr[\Gamma \dif M ] = 0
        \label{eq:PhaseShiftCond}
    \end{align}
    Now, by the chain rule,
	\begin{align}
		\dif g = \sum_i \dif \Psi_i  \frac{\partial g}{\partial \Psi_i} +  \dif \Psi_i^*  \frac{\partial g}{\partial \Psi_i^*}
	\end{align}
	where we treat each component of the wavefunction $\Psi_i = \braket{i|\Psi}$ and its complex conjugate as independent variables (we have also temporarily suppressed the $O$-dependence of $\ket{\Psi}$).

 The modulus of the above can be bounded using the Cauchy-Schwarz inequality
	\begin{align}
		|\dif g| \leq 2\|\dif \Psi \| \|\partial_\Psi g\| , 
		\label{eq:DifGChain}
	\end{align}
	where we use the shorthand $\partial_\Psi g  = \sum_i\bra{i} \partial_{\Psi_i}g$ and $\| v \| = \sqrt{\sum_i |v_i|^2}$ is the vector norm. In terms of the Lie algebra element $\dif A$, the change in the many-body state is
	\begin{align}
		\ket{\dif \Psi} = \frac{1}{4}\sum_{jk} \gamma_j [\dif M]_{jk}\gamma_k\ket{\Psi} + \iu \dif \theta \ket{\Psi}
	\end{align}
	We choose $\dif \theta = \frac{\iu}{4} \sum_{jk} [\dif M]_{jk} \braket{\Psi|\gamma_j\gamma_k|\Psi} = -\frac{1}{4}\Tr[\Gamma \dif A]$, such that $\braket{\Psi|\dif \Psi} = 0$. The magnitude satisfies
	\begin{align}
		\braket{\dif \Psi|\dif \Psi} 
            &= \frac{1}{16}\sum_{ijkl} [\dif M]_{ji} [\dif M]_{kl} \braket{\Psi|\gamma_i\gamma_j \gamma_{k}\gamma_{l}|\Psi} - \frac{1}{16}\Tr[\Gamma \dif M]^2 \nonumber\\
            &= \frac{1}{16}\sum_{ijkl} [\dif M]_{ji} [\dif M]_{kl} \Big( \langle \gamma_i \gamma_j \rangle \langle \gamma_k \gamma_l \rangle - \langle \gamma_i \gamma_k \rangle \langle \gamma_j \gamma_l \rangle + \langle \gamma_i \gamma_l \rangle \langle \gamma_j \gamma_k \rangle\Big) - \frac{1}{16}\Tr[\Gamma \dif M]^2 \nonumber\\
            &= \frac{1}{8} \left( \| \dif M \|^2_2 - \Tr [\Gamma (\dif M) \Gamma (\dif M)] \right) \leq \frac{1}{8} \left( \| \dif M \|^2_2 + \left\vert \Tr [\Gamma (\dif M) \Gamma (\dif M)] \right\vert \right)\nonumber\\
            &\leq \frac{1}{8} \left( \| \dif M \|^2_2 + \| \Gamma \dif M \|^2_2 \right) \nonumber\\
            &\leq \frac{\| \dif M \|^2_2}{8} \left( 1+\| \Gamma \|^2_\infty \right) \leq \frac{1}{4}\|\dif M\|_2^2,
		\label{eq:DifPsi}
	\end{align}
	where the second line follows from Wick's theorem for the Gaussian state $\ket{\Psi}$, the third equality follows from $\langle \gamma_j \gamma_k\rangle = -i \Gamma_{jk} + \delta_{jk}$,
 the fourth line follows from the Cauchy-Schwarz inequality for the Hilbert-Schmidt inner product $\vert \Tr [AB]\vert \leq \|A\|_2 \|B\|_2$, and the last step is H{\"o}lder's inequality. By combining Eqs.~\eqref{eq:DifGChain} and \eqref{eq:DifPsi} with Eq.~\eqref{eq:DifODA}, from the definition of the Lipschitz constant $\eta$ [Eq.~\eqref{eq:Lipschitz}] we get
	\begin{align}
		\eta \leq \sup_{O} \|\partial_{\Psi} g(\Psi(O))\|%
		\label{eq:LipschitzPsiDeriv}
	\end{align}
	All that remains is to bound $\partial_{\Psi} g(\Psi)$. Differentiating Eq.~\eqref{eq:GODef} with respect to the components of the state $\ket{\Psi}$, we can identify two separate contributions $\frac{\partial g}{\partial \Psi_i} = \braket{v|i} + \braket{w|i}$, where
	\begin{align}
	    \bra{v} &=  
		\bra{\Psi}\sum_\mathbf{m} f(\Gamma^A_\mathbf{m}) 
    \big(\mathbbm{1}_A \otimes P_\mathbf{m} \big) \label{eq:diffGvDef} \\  %
		\bra{w} &= \bra{\Psi}\sum_{\mathbf{m}} \sum_{jk}  \left[\frac{\partial f(\Gamma^A_{\mathbf{m}})}{\partial [\Gamma^A_\mathbf{m}]_{jk}} \right]
  \Big((\pi_{jk} - [\Gamma^A_\mathbf{m}]_{jk})\otimes P_\mathbf{m} \Big),\label{eq:diffGwDef}
	\end{align}where $P_\mathbf{m} = \ketbra{\mathbf{m}}{\mathbf{m}}_B$. By the triangle inequality, $\|\partial_\Psi g\| \leq \|v\| + \|w\|$. Firstly, we can evaluate
	\begin{align}
		\|w\|^2 &= 
            \sum_{\mathbf{m}} \sum_{ijkl}
		\braket{\Psi| (\pi_{ij} - [\Gamma^A_\mathbf{m}]_{ij})(\pi_{kl} - [\Gamma^A_\mathbf{m}]_{kl}) \otimes P_\mathbf{m} |\Psi} \left[\frac{\partial f(\Gamma^A_{\mathbf{m}})}{\partial [\Gamma^A_\mathbf{m}]_{ij}} \frac{\partial f(\Gamma^A_{\mathbf{m}})}{\partial [\Gamma^A_\mathbf{m}]_{kl}} \right]  \nonumber\\
        &=
        \sum_\mathbf{m} p_\mathbf{m} \sum_{ijkl} \frac{\partial f(\Gamma^A_{\mathbf{m}})}{\partial [\Gamma^A_\mathbf{m}]_{ij}} \frac{\partial f(\Gamma^A_{\mathbf{m}})}{\partial [\Gamma^A_\mathbf{m}]_{kl}} \Big( - [\Gamma^A_\mathbf{m}]_{ik}[\Gamma^A_\mathbf{m}]_{jl} + [\Gamma^A_\mathbf{m}]_{il}[\Gamma^A_\mathbf{m}]_{jk} + \delta_{ik}\delta_{jl} - \delta_{il}\delta_{jk}\Big)
        \nonumber\\
                    &\leq \sup_{\Gamma^A \in \Upsilon^A}\Big(- 2\Tr\big[\Gamma^A (\nabla f)\Gamma^A (\nabla f)  \big] + 2\Tr\big[(\nabla f)^2\big]\Big) \nonumber\\
                    &\leq
            \sup_{\Gamma^A \in \Upsilon^A} \|\nabla f\|_2^2(2 \|\Gamma^A\|_\infty^2 + 2) \leq 4\|\nabla f\|_2^2.
	\end{align}
	In moving from the first to the second line in the above, we have used the fact that $\braket{\Psi|X_A \otimes P_\mathbf{m}|\Psi} = p_\mathbf{m}\braket{\phi_A(\mathbf{m})|X_A|\phi_A(\mathbf{m})}$ for any operator $X_A$, along with Wick's theorem and the expression $\braket{\phi_A(\mathbf{m})|\iu \gamma_j \gamma_k|\phi_A(\mathbf{m})} = \Gamma_{jk} + \iu \delta_{jk}$. %
	Secondly, we clearly have
 \begin{align}
     \|v\|^2 = \sum_\mathbf{m} p_\mathbf{m} f(\Gamma^A_\mathbf{m})^2 \leq \left(\sup_{\Gamma^A \in \Upsilon^A} f(\Gamma)\right)^2
 \end{align}
	Overall, this gives us $\|\partial_\Psi g\| \leq 2 \|\nabla f\|_2 + \sup_{\Gamma^A} |f(\Gamma^A)|$. From our assumptions on $f$, we have $\|\nabla f\|_2 \leq c$ and $\sup_{\Gamma^A}|f(\Gamma^A)| \leq \frac{1}{2}(b-a)$, so $\|\partial_\Psi g\| \leq 2c + \frac{b-a}{2}$. When combined with Lemma \ref{lemma:Levy} for the case $n = 2L = 2(L_A+L_B)$ and using Eq.~\eqref{eq:LipschitzPsiDeriv}, this gives
	\begin{align}
		\mathbbm{P}_{O \sim \textrm{Haar}} \left( |g(O) - \braket{g}_{\rm Haar}| > \epsilon \right) \leq 2 
        \exp\left(-\frac{(L_B+L_A-1)\epsilon^2}{ 16[c + (b-a)/4]^2}\right)
	\end{align}
	Together with Eq.~\eqref{eq:PEHaarAvg} and using $L_B+L_A - 1 \geq L_B$, we arrive at Eq.~\eqref{eq:FConcentration}. \hfill $\square$

\section{Wasserstein-1 distance details and upper bound \label{app:Thm2proof}}

In this appendix, we present our proof of Theorem~\ref{thm:Wasserstein}, which is an upper bound on the Wasserstein-1 distance between the projected ensemble and the Gaussian Haar ensemble, averaged over globally random pre-measurement states. We also review a number of other well-known distance measures between probability distributions, and discuss their appropriateness for quantum information problems of the kind studied in this work.

\subsection{Definition and properties of Wasserstein-1 distance \label{app:WassersteinDiscussion}}

The Wasserstein-1 distance was introduced in the main text, but here we give a more full definition for completeness. Given two probability measures
$\mu, \nu$ over a space $\mathcal{M}$ endowed with a metric $d$, the Wasserstein-$p$ distance for $p \in [1,\infty)$ is defined as
\begin{align}
    W_p(\mu, \nu) = \inf_{\gamma \in \Pi(\mu, \nu)} \left(\mathbbm{E}_{(x,y) \sim \gamma}  d(x,y)^p \right)^{1/p}
\end{align}
where the set $\Pi(\mu, \nu)$ denotes all \textit{couplings} of $\mu$ with $\nu$, i.e.~probability measures  $\gamma$
over $\mathcal{M}\times \mathcal{M} \ni (x,y)$ such that the marginals on each copy of $\mathcal{M}$ are $\mu$ and $\nu$, respectively. %
This is a highly general definition, and is manifestly related to the mathematical problem of optimal transport: loosely speaking, the distribution $\gamma$ that achieves the infinum represents the most efficient way to `move' probability mass between various regions of $\mathcal{M}$, where efficiency is quantified in terms of the distance that each mass element is moved \cite{Villani2009}. 

For our purpose,
we will focus specifically on the case $p = 1$, in which case it will be more convenient to use the Kantorovich-Rubinstein dual characterization of the Wasserstein metric \cite{Kantorovich1958}
\begin{align}
    W_1(\mu, \nu) = \sup_{\substack{f : \mathcal{M} \rightarrow \mathbbm{R}\\ \textrm{Lip}(f) \leq 1}} \mu(f) - \nu(f),
    \label{eq:W1Dual}
\end{align}
where the supremum is over all functions $f$ whose Lipschitz constant $\text{Lip}(f) = \sup_{x,y} |f(x) - f(y)|/d(x,y)$ is no greater than unity. Here and throughout, we use the shorthand for expectation values of functions $\mu(f) \coloneqq \int_{\mathcal{M}} \dif \mu(x) f(x)$, where $\dif \mu(x)$ is the density corresponding to the measure $\mu$.

This alternative expression allows a more natural comparison to a wider class of distances between probability distributions, known collectively as \textit{integral probability metrics} (IPMs).
Given some class of real-valued functions $\mathcal{F}$, we define the corresponding IPM as
\begin{align}
    \text{IPM}_{\mathcal{F}}(\mu, \nu) \coloneqq \sup_{\substack{f : \mathcal{M} \rightarrow \mathbbm{R}\\ f \in \mathcal{F}}} \mu(f) - \nu(f).
\end{align}
For example, the distance $\Delta^{(k)}$ defined via the trace distance in Eq.~\eqref{eq:DeltaKDef} can be written in the above form if one takes $\mathcal{F}$ to be the class of degree-$k$ polynomials with appropriately bounded norm. Another example of an IPM is the total variation distance (TVD), for which the class $\mathcal{F}_{\rm TVD} = \{f : \mathcal{M} \rightarrow [0,1]\}$ is the set of all functions that evaluate to points on the unit interval. The Kullback-Leibler divergence also has an analogous dual expression in the form $D_{\rm KL}(\mu | \nu) = \sup\{\mu(f) : \nu(e^f) = 1\}$ \cite{Gray2011}, although this is not strictly an IPM.

The significance of polynomial moments and the Wasserstein-1 distance is that the class of `test functions' $\mathcal{F}$ in these cases incorporates some notion of smoothness. In contrast, the dual representations of TVD and KL divergence include test functions that can be arbitrarily discontinuous. This is why the latter two distance measures always give vacuous results when comparing continuous and discrete distributions: for example, the test functions can be chosen such that they can distinguish between a perfect delta function versus an arbitrarily narrow Gaussian.

We now turn back to the problem of comparing the projected ensemble and the Gaussian Haar ensemble, which in our setting are distributions over covariance matrices $\Gamma$ of pure Gaussian states. The space $\mathcal{M}$ is this case should be thought of as the space of valid covariance matrices $\Upsilon_{\rm pure}$. Writing $ \nu_{\rm PE}(\Gamma)$ and $\nu_{\rm GHE}(\Gamma)$ as the probability measures of the corresponding ensembles over $\mathcal{M}$, we suggest that the Wasserstein-1 distance
\begin{align}
    W_1( \nu_{\rm PE}, \nu_{\rm GHE})
\end{align}
is an appropriate measure of deep thermalization. We emphasize that, unlike many other statistical distances between probability distributions, the Wasserstein-1 distance can quantitatively compare the discrete projected ensemble with the continuous Haar ensemble.

\subsection{Proof of Theorem~\ref{thm:Wasserstein}: upper bounding the Wasserstein-1 distance}

We now move on to the main proof, which pertains to the distribution of a specific scalar function $\theta$ of the 
post-measurement covariance matrix $\Gamma^A_\mathbf{m}$%
---this contains less information than the multivariate
distribution over the full matrix $\Gamma^A_\mathbf{m}$, but given that we have freedom in choosing the function $\theta(\Gamma^A_\mathbf{m})$, this is still a powerful characterization. For notational convenience, we will just use $\theta$ to denote the random variable $\theta(\Gamma^A)$, where $\Gamma^A$ is distributed according to either the projected or Gaussian Haar ensembles. The distributions of $\theta$ in each ensemble we write as  $\dif \nu_\theta(\theta)$ and $\dif \mu_\theta(\theta)$, respectively. %
Our aim is to derive an upper bound on the  distance
\begin{align}
    \Delta_{\theta} \coloneqq \mathbbm{E}_{\Psi} W_1(\nu_\theta, \mu_\theta),
\end{align}
averaged over Gaussian pre-measurement states $\ket{\Psi}$. For example, $\theta(\Gamma^A_\mathbf{m})$ might be the purity of some subregion $A_1 \subset A$, in which case a small value of $\Delta_{\theta}$ would imply that the full distribution of purities in the projected ensemble is close to that of the Gaussian Haar ensemble.

As described in the main text, we put some restrictions on the function $\theta$. Firstly, its value should be bounded; by shifting by a constant and rescaling, we can assume without loss of generality that
\begin{align}
    \theta(\Gamma^A) &\in [-1,1], \;\; \forall \textrm{ CM }\Gamma^A \in \Upsilon^A.
\end{align}
Secondly, we assert that $\theta$ is differentiable, and is sufficiently smooth, in the sense that it is $c$-Lipschitz
\begin{align}
    \sum_{ij} \left|\frac{\partial \theta(\Gamma^A)}{\partial [\Gamma^A]_{ij}}\right|^2 \leq c^2
    \label{eq:FSmooth}
\end{align}
for some constant $c$.  %

The high-level structure of our argument goes as follows: Suppose that $W_1(\nu_\theta, \mu_\theta) > \omega$ 
for some $\omega > 0$. Then, by Eq.~(\ref{eq:W1Dual}), there would exist a 1-Lipschitz test function $f(\theta)$ for which $\nu_{\theta}(f) - \mu_{\theta}(f) > \omega$, which in turn implies that $\nu_{\rm PE}(g) - \nu_{\rm GHE}(g) > \omega$, where $g = f \circ \theta$ is a composition of functions, and is $c$-Lipschitz. However, Lipschitz functions can be well-approximated by degree-$k$ polynomials with an error decreasing as $k$ increases. Using Lemma \ref{lemma:FConcentration}, we infer that the averages of such polynomials are very close to their corresponding values in the GHE for large enough $L_B$, and together this tells us that $\omega$ cannot be too large. The main subtlety we have to overcome is the fact that the average over $\Psi$ and the supremum in Eq.~\eqref{eq:W1Dual} do not commute; that is, for each random choice of pre-measurement state $\Psi$ we choose a different test function $f$.
Hence, we will need to prove that for a typical instance of $\Psi$, the averages of \textit{all} degree-$k$ polynomials agree in the projected and Haar ensembles, as opposed to just some fixed instance~$f$.

We start with a lemma concerning the approximation result alluded to above.
\begin{lemma}\label{lemma:PolyApprox}
    Let $f: [-1,1] \rightarrow \mathbbm{R}$ be a 1-Lipschitz function, and $k \in \mathbbm{N}$, then there exists a degree-$k$ polynomial
    \begin{align}
        f_k(\theta) = \sum_{r=0}^k a_{r,k}T_r(\theta),
        \label{eq:PolyChebyshev}
    \end{align}
    where $T_r(\theta) = \cos(r \arccos \theta)$ are Chebyshev polynomials of the first kind, that approximates $f$ in the sense that
    \begin{align}
        \sup_{\theta \in [-1,1]} |f(\theta) - f_k(\theta)| \leq \frac{6}{k}.
        \label{eq:PolyError}
    \end{align}
    Moreover, the coefficients in $f_k$ satisfy, for $r \geq 1$,
    \begin{align}
    \label{eq:PolyCoefficientBound}
        |a_{r,k}| &\leq \frac{2}{r}.%
    \end{align}
\end{lemma}
\noindent \textit{Proof of Lemma~\ref{lemma:PolyApprox}.---}The main part of this lemma is a standard result in approximation theory due to Jackson~\cite{Jackson1912}, but since the final statement pertains to the structure of the approximating polynomial itself, here we will describe how $f_k(\theta)$ is constructed, basing our approach on that of Ref.~\onlinecite{Carothers2009}.

First we write $\theta = \cos s$ for $s \in [0,\pi)$, which gives us a well-defined function $h(s) = f(\cos s)$ on this domain. We then extend the domain of $h$ symmetrically to obtain an even periodic function on the unit circle $S^1$.  Since $|\frac{\dif }{\dif s} \cos s| \leq 1$, we have that if $f(\theta)$ is 1-Lipschitz, then so is $h(s)$.
Moreover, degree-$k$ polynomials $f_k$ on $[-1,1]$ translate to degree-$k$ trigonometric polynomials on~$S^1$, namely functions of the form $\frac{1}{2}a_0 + \sum_{r=1}^k a_r \cos(r s)$. Since we can always shift $f(\theta)$ and $f_k(\theta)$ by a constant without changing the degree of approximation, we can assume that $\int_{-\pi}^\pi \dif s \, h(s) = 0$ without loss of generality, and take $a_0 = 0$.

An important object will be the Jackson kernel $J_k(t)$, which is a particular degree-$k$ trigonometric polynomial that is non-negative everywhere and normalized to $\int_{-\pi}^\pi\dif t J_k(t) = 1$. Its particular form (see Ref.~\onlinecite{Jackson1912}) is not important, but it is constructed to be tightly concentrated around zero for large $k$. In terms of this kernel, we can construct a particular polynomial approximation to $h(s)$,
\begin{align}
    h_k(s) &= \int_{-\pi}^\pi \dif t \, J_k(t) h(s + t).
\end{align}
The fact that $h_k(s)$ is a trigonometric polynomial of order $k$ follows from shifting the integration variable $t \rightarrow t - \theta$, and using fact that $J_k(t)$ is a trigonometric polynomial of order $k$ in $t$, we have that by angle addition formulae, $J_k(t-s)$ is a trigonometric polynomial in $s$. The error at a given point $s$ can be written as
\begin{align}
    |h_k(s) - h(s)| &= \left|\int_{-\pi}^{\pi}\dif t \, J_k(t) \Big(h(s+t)-h(s)\Big) \right| \nonumber\\
    &\leq \int_{-\pi}^{\pi}\dif t\, J_k(t) \Big|h(s+t)-h(s)\Big| \nonumber\\
    &\leq \int_{-\pi}^{\pi}\dif t\, |t| J_k(t)  \leq \frac{6}{k}.
\end{align}
In going from the second to the third line, we use the fact that $h(s)$ is 1-Lipschitz, while the final inequality is a particular property of the Jackson kernel. Since the above holds for all 1-Lipschitz $f$ and any $s$, we can convert back from $h_k(\theta)$ to $f_k(\theta) = h_k(\arccos \theta)$, which can be written in terms of Chebyshev polynomials of order up to $k$. This proves Eq.~\eqref{eq:PolyError}.

Using the orthogonality of trigonometric functions, we can obtain an explicit expression for the coefficients $a_{r,k}$ in Eq.~\eqref{eq:PolyChebyshev}, and in turn upper bound their absolute values
\begin{align}
    |a_{r,k}| &=  \left| \frac{1}{\pi} \int_{-\pi}^\pi\dif s \int_{-\pi}^\pi \dif t\, h(t+s)J_k(t)\cos(rs)\right| \nonumber\\ 
    &\leq \left| \frac{1}{2\pi} \int_{-\pi}^\pi\dif s \, \int_{-\pi}^\pi \dif t \, [h(t+\pi/r + s) - h(t+s)]J_k(t)\cos(rs) \right| \nonumber\\
	&\leq \frac{1}{2\pi}\frac{\pi}{r}\int_{-\pi}^\pi\dif t \,J_k(t) \int_{-\pi}^\pi \dif s\, |\cos(rs)| =  \frac{2}{r},%
\end{align}
where we have used the relation $\cos(r[t+\pi/r]) = -\cos(rt)$, and again invoked the fact that $h(t)$ is 1-Lipschitz. \hfill $\square$

Lemma \ref{lemma:PolyApprox} identifies $T_r(\theta) \equiv T_r(\theta(\Gamma))$ as a basis of functions which effectively control the behavior of 1-Lipschitz functions. Using Eq.~\eqref{eq:FSmooth} and the bound $|\partial_\theta T_r(\theta)| \leq r^2$, we can verify that the function $T_r\circ \theta$ satisfies the conditions of Lemma \ref{lemma:FConcentration}, namely it is $cr^2$-Lipschitz, and bounded in $[-1,1]$. Therefore, by Eq.~\eqref{eq:FConcentration}, we have, for any $\epsilon> 0$,
\begin{align}
    \mathbbm{P}_{\Psi}\Big( \left|\nu_{\theta}(T_r) - \mu_{\theta}(T_r)\right| > \epsilon\Big) &\leq 2\exp\left(-\kappa_\theta L_B \epsilon^2\right) & \text{with }\kappa_\theta = \frac{1}{16(cr^2 + 1/2)^2}.
    \label{eq:ChebyshevConcentration}
\end{align}%

Now, for a given choice of $\Psi$ (which determines $\nu_{\rm PE}$, and in turn $\nu_\theta$), if $W_1(\nu_\theta, \mu_\theta) \geq \omega$, then we will be able to find some 1-Lipschitz function $f(\theta)$ such that $\nu_\theta(f) - \mu_\theta(f) > \omega$. Any such function $f(\theta)$ can be approximated by Chebyshev polynomials in the sense of Lemma \ref{lemma:PolyApprox}, and we define the approximating polynomial $f_k(\theta)$ as in Eq.~\eqref{eq:PolyChebyshev} for any  integer $k \geq 1$. %
We shall next use this approximation $f_k(\theta)$ and Eq.~\eqref{eq:ChebyshevConcentration} to bound the probability that $\nu_\theta(f) - \mu_\theta(f) > \omega$.
First, note that, for all 1-Lipschitz functions $f(\theta)$, we have
\begin{align}
    \nu_\theta(f) - \mu_\theta(f) &= \big[\nu_\theta(f) - \nu_\theta(f_k)\big] - \big[\mu_\theta(f) - \mu_\theta(f_k)\big] + \big[\nu_\theta(f_k) - \mu_\theta(f_k)\big] \nonumber\\
    &\leq \frac{12}{k} + \big[\nu_\theta(f_k) - \mu_\theta(f_k)] \nonumber\\
    &\leq \frac{12}{k} + \sum_{r=1}^k |a_{k,r}|\Big|\nu_\theta(T_r) - \mu_\theta(T_r)\Big| \nonumber\\
    &\leq \frac{12}{k} + \sum_{r=1}^k \frac{2}{r}
    \Big|\nu_\theta(T_r) - \mu_\theta(T_r)\Big|,
    \label{eq:FDifference}
\end{align}where we used Eq.~\eqref{eq:PolyError} twice to obtain the first inequality. Then, let $\{\alpha_r\}_{r=1}^k$ be an arbitrary collection of non-negative numbers satisfying $\sum_{r=1}^k \alpha_r \leq 1$. For some $\omega > 12/k$, if we have $\nu_\theta(f) - \mu_\theta(f) > \omega$, then from \eqref{eq:FDifference} there must be at least one $r$ for which
\begin{align}
    \frac{2}{r}\Big|\nu_\theta(T_r) - \mu_\theta(T_r)\Big|  > \left(\omega - \frac{12}{k}\right) \alpha_r.
\end{align}
The probability of this occurring can be bounded using Eq.~\eqref{eq:ChebyshevConcentration}, along with Boole's inequality, giving
\begin{align}
    \mathbbm{P}_\Psi\left(\exists f(\theta)\,\text{1-Lipschitz s.t. }\nu_\theta(f) - \mu_\theta(f) > \omega \right) &\leq \mathbbm{P}_\Psi\left(\bigcup_{r=1}^k \; \frac{2}{r}\Big|\nu_\theta(T_r) - \mu_\theta(T_r)\Big|  > \left(\omega - \frac{12}{k}\right) \alpha_r \right) \nonumber\\  %
    &\leq 2\sum_{r=1}^k \exp\left( - \frac{L_B \alpha_r^2 r^2(\omega - 12/k)^2}{64(r^2c + 1/2)^2} \right).
\end{align}
Now, if $\theta$ is $c'$-Lipschitz with $c' < 1/2$, then it is trivially also $c$-Lipschitz with $c \geq 1/2$. Thus we can assume $c \geq 1/2$ here, and hence $(r^2c + 1/2) \leq c(r^2+1) \leq 2cr^2$. We then set $\alpha_r = r/k^2$, and from the dual definition of the Wasserstein distance \eqref{eq:W1Dual} we get
\begin{align}
    \mathbbm{P}_{\Psi}\Big(W_1(\nu_\theta, \mu_\theta) > \omega \Big) \leq 2k \exp\left( - \frac{L_B(\omega - 12/k)^2}{256c^2k^4}\right), \label{eq:prW1bound}
\end{align}
for any positive integer $k > 12/\omega$. We can translate this into a bound on the mean using $\mathbbm{E}_{\Psi}[W_1] = \int_0^\infty \dif \omega \, \mathbbm{P}(W_1 > \omega)$.
Using that Eq.~\eqref{eq:prW1bound} is valid for $\omega > 12/k$, we thus have
\begin{align}
    \mathbbm{E}_{\Psi} \left[W_1(\nu_\theta, \mu_\theta)\right] &= 
    \int_0^{\frac{12}{k}} \dif \omega\, \mathbbm{P}(W_1 > \omega) + \int_0^\infty \dif \epsilon \, \min\left(1, 2k \exp\left( - \frac{L_B \epsilon^2}{256 c^2 k^4 }\right) \right) \\
    &\leq \frac{12}{k} + \int_0^\infty \dif \epsilon \, \min\left(1, 2k \exp\left( - \frac{\kappa L_B \epsilon^2}{256 c^2 k^4}\right) \right) \\
    &\leq \frac{12}{k} + \frac{16ck^2}{\sqrt{ L_B}} \left( 1+ \sqrt{\log 2k}\right).
\end{align}%
We see that a near-optimal choice for the degree of approximation is $k^3 = 3\sqrt{L_B}/4c$.
Since we have taken $c \geq 1/2$, then assuming $L_B \geq 2$, we have $\log(2k) \leq \frac{1}{6}\log(2^6 \cdot (3/4)^2 \cdot 2^{-2} \cdot L_B) \leq \frac{1}{6}(\log L_B + 2 \log 3) \leq \frac{1}{6}(1 + \frac{2\log3}{\log 2})\log L_B \leq  \log L_B$. 
This gives us a decay of the Wasserstein distance as
\begin{align}
    \mathbbm{E}_{\Psi} W_1(\nu_\theta, \mu_\theta) \leq \frac{\alpha\sqrt{\log L_B}}{L_B^{1/6}}, 
\end{align}where the  constant of proportionality can be bounded as $\alpha \leq c^{1/3}(12 \cdot (4/3)^{1/3} + 32 \cdot (3/4)^{2/3}) \leq 40 c^{1/3}$.
Recalling that we made an assumption that $c \geq 1/2$ earlier, to include more general values of $c$ we can set $\alpha = 40 \max(c^{1/3}, 1)$.
If $\theta$ is $c$-Lipschitz and bounded in a different interval $[a,b]$, then the function $\theta' = \frac{2}{b-a}(\theta + \frac{a+b}{2})$ is bounded in $[-1,1]$ and is $c' = 2c/(b-a)$-Lipschitz. From the exact expression for the Wasserstein-1 distance for one-dimensional distributions, Eq.~\eqref{eq:W11D}, we have that $W_1(\nu_\theta, \mu_\theta) = \frac{1}{2}(b-a)W_1(\nu_{\theta'}, \mu_{\theta'})$, and so a valid choice of the constant is
\begin{align}
    \alpha = 40\max\left(\frac{1}{2^{2/3}}c^{1/3}(b-a)^{2/3}, \frac{1}{2}(b-a)\right) \leq 27 (b-a)^{2/3} \max(c^{1/3}, (b-a)^{1/3})
\end{align}
This concludes our proof of Theorem~\ref{thm:Wasserstein}.\hfill $\square$

\subsection{Examples of physical quantities}

Theorem \ref{thm:Wasserstein} puts some conditions on the quantities $\theta(\Gamma^A)$ whose distribution we are considering---specifically, the quantities must be bounded in some interval, and also have bounded derivatives $\|\nabla \theta\|_2 \leq c$. Here we briefly describe three examples of physical significance that satisfy these conditions.

\subsubsection{$n$-point correlation functions}

Firstly, from Eq.~\eqref{eq:PfaffianInverse} and the discussion immediately afterwards, we already have that for $n$ even, any $n$-point correlation function (with $n$ even) can be written as a Pfaffian of the covariance matrix $\braket{\phi_A(\mathbf{m})|\gamma_{j_1} \cdots \gamma_{j_n}|\phi_A(\mathbf{m})} = \Pf(\Gamma|_S)$, where $S = \{j_1, \ldots, j_n\} \subseteq [2L_A]$ is the set of indices of the Majorana operators included in the product. Because $\|\gamma_{j_1} \cdots \gamma_{j_n}\|_\infty = 1$, such a correlator is bounded in $[-1,1]$, and we have already shown that $c = \|\nabla \Pf(\Gamma^A_S)\|_2 \leq \sqrt{2L_A}$, hence Theorem \ref{thm:Wasserstein} applies.

\subsubsection{Subsystem R\'enyi entropies}
Secondly, we consider subsystem $\alpha$-R\'enyi entropies. Recall the definition of R\'enyi entropies
\begin{align}
    S^{(\alpha)}(\rho) = \frac{1}{1-\alpha}\log\big( \Tr[\rho^\alpha]\big)
\end{align}
which generalize the von Neumann entropy $S(\rho)$ in the sense that $\lim_{\alpha \rightarrow 1} S^{(\alpha)}(\rho) = S(\rho)$. In the case where $\rho = \rho(\Gamma)$ is a Gaussian fermionic state with covariance matrix $\Gamma$, whose eigenvalues we write as $\pm \iu s_i$, we can express the R{\'e}nyi entropies as
\begin{align}
    S^{(\alpha)}\big(\rho(\Gamma)\big) = \sum_i \frac{1}{1-\alpha}\log\left[\left(\frac{1+s_i}{2}\right)^\alpha + \left(\frac{1-s_i}{2}\right)^\alpha\right],
\end{align}
which is the natural analogue of Eq.~\eqref{eq:vonNeumannSV} (here the sum over $i$ is over each pair of eigenvalues). We can apply this to the case where $\rho$ is the reduced state on some subsystem $A_1 \subseteq A$, in which case we have $\Gamma = \Gamma^{A_1}_\mathbf{m}$. These quantities are all bounded in $[0, L_{A_1}\log 2]$, and also can be shown to have bounded derivatives: The derivatives of the eigenvalues of $\Gamma$ are themselves bounded $\|\nabla_\Gamma s_i\|_2 \leq 1$, and for $\alpha \in (1,\infty)$ we have
\begin{align}
    |\partial_{s_i} S^{(\alpha)}| = \left|\frac{\alpha}{1-\alpha}\frac{(\frac{1+s_i}{2})^{\alpha-1} - (\frac{1-s_i}{2})^{\alpha-1}}{(\frac{1+s_i}{2})^{\alpha} + (\frac{1-s_i}{2})^{\alpha}}\right| \leq \frac{2\alpha}{\alpha-1}
\end{align}
Thus, viewed as a function of $\Gamma^A$, the subsystem R\'enyi entropy $S^{(\alpha)}(\rho^{A_1})$ satisfies the conditions of Thm~\ref{thm:Wasserstein} with $[a,b] = [0,L_A \log 2]$ and $c \leq 2L_{A_1}\alpha/(\alpha-1)$ for $\alpha > 1$. Hence the coefficient $\kappa$ appearing in Eq.~\eqref{eq:W1Decay} can be taken to be $\kappa = 54 L_{A_1} \frac{\alpha}{\alpha-1}$, which is finite as long as we are away from the von Neumann limit $\alpha \rightarrow 1$. We provide an alternative proof for the $\alpha=1$ case next.

\subsubsection{Subsystem von Neumann entropy}
Thirdly, we consider the subsystem von Neumann entropy.
For a pure (not necessarily Gaussian) many-body state $\ket{\Psi}$ and a subsystem $A$ with reduced density matrix $\rho^A = \Tr_B [\ketbra{\Psi}{\Psi}]$, we denote the subsystem von Neumann entropy, viewed as a function of $\ket{\Psi}$, by
\begin{align}
    G(\ket{\Psi}) \coloneq S(\rho^A) = - \Tr \left[\rho^A \log \rho^A\right]. \label{eq:G_Lipschitz}
\end{align}%
Using Lemma III.2 in Ref.~\onlinecite{Hayden_2006}, we have that, for any two pure states $\ket{\Psi}$ and $ \ket{\Phi}$, $G$ is $\eta_G$-Lipschitz
\begin{align}
    &| G(\ket{\Psi}) - G(\ket{\Phi})| \leq \eta_G \| \ket{\Psi}- \ket{\Phi}\|, &  \eta_G \leq 2\sqrt{2}L_A. 
\end{align}

We now turn to Gaussian many-body states $\ket{\Psi(O)}$ parametrized by rotations $O \in \mathrm{SO}(2L)$. Here, we denote the subsystem von Neumann entropy, viewed as a function of $O$, by 
\begin{align}
    F(O) \coloneq G(\ket{\Psi(O)}).
\end{align}Since $F$ is bounded in the interval $[0, L_A]$, it remains to show that it is $c$-Lipschitz. To do so, consider two infinitessimally close states induced by the transformation $O \to e^{\dif M} O$, as per App.~\ref{app:lemma1_proof},
\begin{align}
    \left|F(e^{\dif M}O) - F(O)\right| = \left| G(\ket{\Psi(e^{\dif M}O)}) - G(\ket{\Psi(O)})\right| %
    \leq \eta_G \left\| \ket{\Psi(e^{\dif M}O)}- \ket{\Psi(O)}\right\| %
    \leq \frac{\eta_G}{2} \| \dif O \|_2,
\end{align}where we applied Eq.~\eqref{eq:G_Lipschitz} for the first and Eq.~\eqref{eq:DifPsi} for the second inequality, respectively. Thus, we see that $F$ is $c$-Lipschitz with $c \leq \sqrt{2} L_A$, and accordingly we can apply Thm.~\ref{thm:Wasserstein}, where the constant from Eq.~\eqref{eq:W1Decay} can be taken to be $\kappa=39 L_A$.

\section{Diffusive spreading of correlations \label{app:diff}}

\begin{figure*}[t]
    \centering
        \includegraphics[width=14cm]{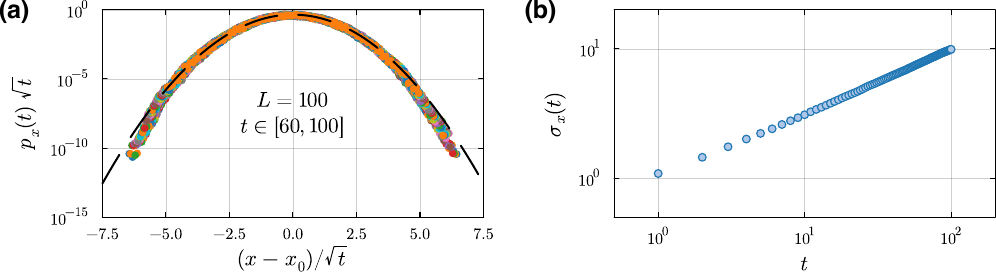}
    \caption{%
             Diffusive spreading of correlations under local matchgates circuits without particle number conservation.
             (a) Probability $p_x(t)$ of a semiclassical wavepacket initialized at $x_0$ to spread at $x$ at time $t$. 
             Different colors correspond to different times $t$, and the data is averaged over $N=10^2$ circuit instances.
             The scaling $p_x(t) \propto  \exp \left[-\frac{(x-x_0)^2}{2t}\right]/\sqrt{t}$ (dashed black line) supports diffusive spreading.   
             (b) Displacement variance $\sigma_x(t)$ vs evolution time $t$. The scaling  $\sigma_x(t)\propto \sqrt{t}$ supports diffusive spreading.
            }
    \label{fig:diff_spreading}
\end{figure*}

In this appendix, we discuss a semiclassical model for the spreading of correlations under local Gaussian dynamics which suggest that this spreading is diffusive. 
We focus on a system of $L$ fermionic modes evolving under 2-local Gaussian unitary gates [Fig.~\ref{fig:setup}(a)]. 

A simple way to characterize the spreading of correlations is through the two-point correlation functions~\cite{diff_spreading}
\begin{align}
    C_i (t) = \frac{1}{\sqrt{2}} \bra{+\cdots +} \gamma_i(t) \gamma_L \ket{+\cdots +}, \label{eq:C_i_t}
\end{align} with $i=1,\dots, 2L$, where $\ket{+ \cdots +} = \ket{+}^{\otimes L}$ is the $+1$ eigenstate of the Pauli operators $X_j$ for $j = 1, \dots, L$ obtained from the Jordan-Wigner transformation 
\begin{align}
    \gamma_{2j-1} =  \prod_{k<j} (-Z_k) X_j , \quad
    \gamma_{2j} =  \prod_{k<j} (-Z_k) Y_j,
\end{align}and $\gamma_i(t) = \sum_{j=1}^{2L} R_{ij}(t) \gamma_j$, where $R(t)$ is the rotation corresponding to the circuit up to time $t$ [App.~\ref{app:Gaussian}]. 
We stress that our dynamics is not particle number conserving, unlike in Ref.~\onlinecite{diff_spreading}, which considers Hamiltonian evolution.

Using these correlation functions, one can define a probability distribution 
\begin{align}
    p_x(t) = \vert C_{2x-1}(t)\vert^2 + \vert C_{2x}(t)\vert^2, \label{eq:p_x_t}
\end{align}for $x = 1, \dots, L$, which satisfies $\sum_{x=1}^L p_x(t) = 1$ for all $t$. This distribution captures if there are any correlations between mode $x$ and the middle modes $L/2$ and $L/2 +1$ at time $t$. A simple calculation at $t=0$ yields $p_x(0) = \frac{1}{2}$ for $x=L/2$ and $L/2+1$, and $p_x(0)=0$ otherwise. Thus, Eq.~(\ref{eq:C_i_t}) characterizes a semiclassical wavepacket initialized at the middle of the chain at time $t=0$ (assuming $L$ is even) and later spreading under the local dynamics.

We can quantify the spreading of correlations by the probability distribution $p_x(t)$ and the variance of the displacement%
\begin{align}
    \sigma^2_x(t) = \sum^L_{x=1} (x-x_0)^2 p_x(t)
\end{align}from the middle of the chain $x_0 = \frac{1}{2}\left[ \frac{L}{2} + \left(\frac{L}{2} +1 \right) \right]$. These are expected to show diffusive behavior, i.e., $\sigma_x(t) \propto \sqrt{t}$ and $p_x(t) \propto \frac{1}{\sqrt{t}} \exp \left[ -\frac{(x-x_0)^2}{2t}\right]$, as in the particle number conserving Hamiltonian dynamics considered in Ref.~\onlinecite{diff_spreading}.

We numerically test these expectations by computing $p_x(t)$ and $\sigma_x(t)$ for a circuit instance and then averaging these over more circuit instances. These computations are efficient, taking poly$(L)$ time since the evolution of $C_i(t)$ reduces to the evolution of $\gamma_i(t)$. We find good agreement between our simulations and these expectations, suggesting diffusive spreading [Fig.~\ref{fig:diff_spreading}(a),(b)].

\section{Additional Wasserstein-1 metric numerics \label{app:extra_W1_numerics}}

\begin{figure*}[t]
    \centering
        \includegraphics[width=17.8cm]{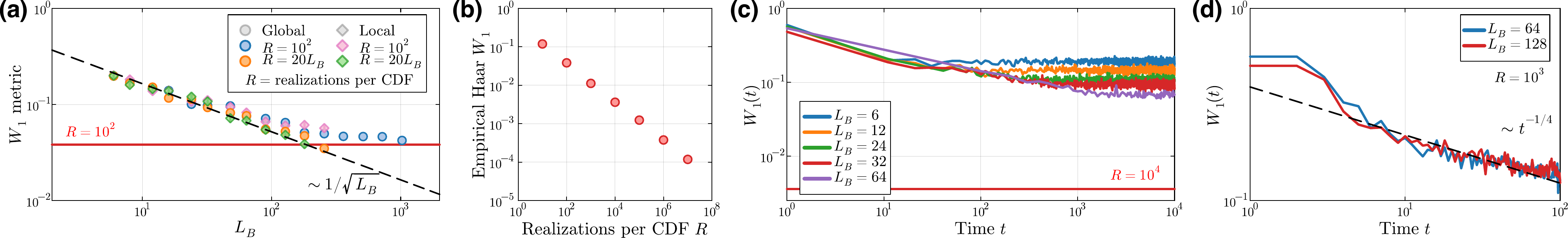}
    \caption{%
             Deep thermalization of global and local fermionic Gaussian dynamics: Wasserstein-1 distance $W_1$ between distributions of a single correlator $\theta(\Gamma^A_{\mathbf{m}}) =\braket{\iu \gamma_1\gamma_2}_{\mathbf{m}}$ for $L_A=6$, averaged over $N=10^2$ circuit instances.
             (a) $W_1$ decays with the number of measured qubits $L_B$ as $W_1 \propto 1/\sqrt{L_B}$ (dashed black line) for both global (circles) and late-time local ($t=10^4$, rhombi) dynamics down to a threshold value set by by the number of realizations per CDF $R$ and shown as the horizontal red line. The error bars, where visible, show the standard error of the mean (SE).
             (b) Estimate of the finite-$R$ plateau from comparing an empirical CDF using $R$ GHE realizations and the exact GHE CDF, showing $W_1 \propto 1/\sqrt{R}$.
             (c) $W_1(t)$ distance vs evolution time $t$ for local dynamics. The horizontal red line shows the empirical threshold. Relative errors are all below $9\%$, and are not shown. An empirical PE is constructed every 10 (100) time steps for $L_B = 6, 12, 24, 32$ ($64$) and $R=10^4$.
             (d) Closer view of the decay $W_1(t) \propto t^{-1/4}$ (dashed black line) for $R=10^3$ and an empirical PE built at each time step. Relative SEs are all below $9\%$, and are not shown.
            }
    \label{fig:app_W1}
\end{figure*}

In this appendix, we provide further numerical evidence for the decay of the Wasserstein-1 metric $W_1$ with the measured subsystem size $L_B$ and time $t$. We use the same setup as in the main text for our simulations.

\subsection{Two-point correlators}
We first focus on the observable $\theta = \braket{\iu \gamma_1 \gamma_2}$. Here, we choose $L_{A_2} = 5$ instead of $L_{A_2}=1$ used in the main text. 
The main features of the behavior of $W_1$ are [compare Fig.~\ref{fig:W1_fig} and Fig.~\ref{fig:app_W1}]: (i) a $W_1(t) \propto t^{-1/4}$ decay down to a value set by $L_B$ for local dynamics, (ii) a decay $W_1 \propto 1/\sqrt{L_B}$ down to a threshold set by the empirical Gaussian Haar ensemble for both the late-time limit of the local dynamics and the global dynamics, and (iii) a central limit theorem like behavior for the decay of the empirical threshold $\propto 1/\sqrt{R}$, where $R$ is the number of sampled states.
The numerical results for different $L_{A_2}$ values are consistent with each other.

We can intuitively understand the dependence of the saturation value on $R$ by considering an empirical CDF $\hat{P}_\mathrm{GHE}$ for the GHE. We construct $\hat{P}^{(R)}_\mathrm{GHE}$ by uniformly sampling $R$ Gaussian states with CM $\gamma^{(i)}_A$ from the GHE, $i = 1, \dots ,R$. As $R$ increases, $\hat{P}_\mathrm{GHE}$ is expected to converge to $P^{(R)}_\mathrm{GHE}$ in a central limit theorem manner: since $\mathbbm{P} \left( W_1 (\hat{P}^{(R)}_\mathrm{GHE}, P_\mathrm{GHE}) > \epsilon \right) < e^{-\Omega(R^2 \epsilon)}$~\cite{Fournier2015}, we have with high probability that $W_1 (\hat{P}^{(R)}_\mathrm{GHE}, P_\mathrm{GHE}) \sim \frac{1}{\sqrt{R}}$;
our numerics agree with this expectation [cf. Figs.~\ref{fig:W1_fig}(b) and~\ref{fig:app_W1}(b)].
If the PE converges to the GHE, i.e., $W_1(P_\mathrm{PE}, P_\mathrm{GHE}) \to 0$ for $L_B \to \infty$, then we intuitively expect that a finite number of realizations
from the PE or GHE are equivalent in the sense of $W_1(\hat{P}^{(R)}_\mathrm{PE}, P_\mathrm{GHE}) \simeq W_1(\hat{P}^{(R)}_\mathrm{GHE}, P_\mathrm{GHE})$; our numerics agree with this expectation [see Figs.~\ref{fig:W1_fig}(a) and~\ref{fig:app_W1}(a)].

In fact, we can make this intuition more rigorous. As discussed in the main text, $\hat{P}^{(R)}_\mathrm{PE}$ for finite $R$ also converges to the full distribution $P_\mathrm{PE}$, that is, $\mathbbm{P} \left( W_1 (\hat{P}^{(R)}_\mathrm{PE}, P_\mathrm{PE}) > \epsilon \right) < e^{-\Omega(R^2 \epsilon)}$; thus, with high probability, $W_1 (\hat{P}^{(R)}_\mathrm{PE}, P_\mathrm{PE}) \sim \frac{1}{\sqrt{R}}$. Applying the triangle and reverse triangle inequality for the Wasserstein-1 metric~\cite{Villani2009}, we have 
\begin{align}
    W_1 (\hat{P}^{(R)}_\mathrm{PE}, P_\mathrm{PE}) - W_1 (P_\mathrm{PE}, P_\mathrm{GHE}) &\leq W_1 (\hat{P}^{(R)}_\mathrm{PE}, P_\mathrm{GHE}) \leq W_1 (\hat{P}^{(R)}_\mathrm{PE}, P_\mathrm{PE}) + W_1 (P_\mathrm{PE}, P_\mathrm{GHE}).
\end{align}%
Taking the $L_B \to \infty$ limit and using $\underset{L_B\to \infty}{\lim} W_1 (P_\mathrm{PE}, P_\mathrm{GHE}) = 0$ (cf. Theorem~\ref{thm:Wasserstein}) yields
\begin{align}
    \frac{\kappa_1}{\sqrt{R}} \leq \lim_{L_B\to \infty} W_1 (\hat{P}^{(R)}_\mathrm{PE}, P_\mathrm{GHE}) \leq \frac{\kappa_2}{\sqrt{R}},
\end{align}for some constants $\kappa_{1,2}$.

\subsection{Entanglement entropies}

We now focus on the subsystem von Neumann and R\'enyi entanglement entropies $\theta = S^{(\alpha)} (\rho^{A_1}_\mathbf{m})$ as our observable choice. Note that the $\alpha =1$ R\'enyi entropy recovers the von Neumann entropy. Here, we choose $L_A = 6$ instead of the $L_A=2$ used in the main text, and consider the entanglement entropies of a half-subsystem with $L_{A_1} = \frac{L_A}{2}$. We also remind the reader that the $W_1$ metric is computed between two empirical distributions, i.e.~$W_1(\hat{P}_\mathrm{PE}, \hat{P}_\mathrm{GHE})$, since the analytical form of the cumulative distributions is not known, to the best of our knowledge; we construct $\hat{P}_\mathrm{GHE}$ by sampling $10^8$ random Gaussian fermionic states.

The main features of the behavior of $W_1$ are [compare Fig.~\ref{fig:W1_entropies} and Fig.~\ref{fig:app_W1_entropies}]: (i) a $W_1(t) \propto 1/\sqrt{t}$ decay down to a value set by $L_B$ for local dynamics and (ii) a decay $W_1(t\gg1) \propto 1/L_B$ for the saturation value.
The numerical results for different $L_A$ values are consistent with each other.

\begin{figure}[t]
    \centering
        \includegraphics[width=8.6cm]{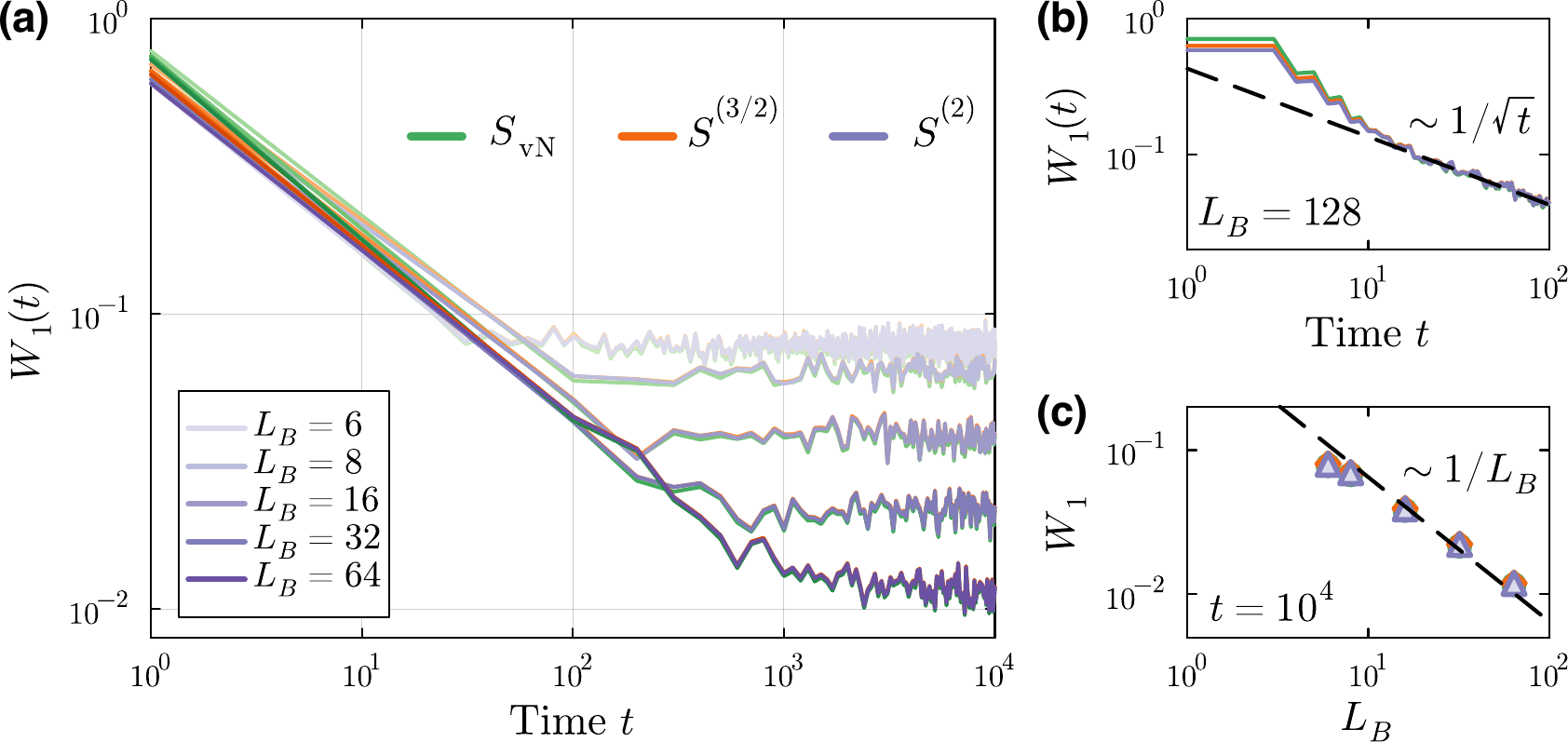}%
    \caption{%
            Deep thermalization of local fermionic Gaussian dynamics: Wasserstein-1 distance $W_1$ between distributions of subsystem von Neumann or R\'enyi entropy $\theta(\Gamma^A_{\mathbf{m}}) = S^{(\alpha)}(\rho^{A_1}_\mathbf{m})$ for $L_A=6, L_{A_1}=\frac{L_A}{2}$ and $N=10^2$ circuit instances.
             (a) $W_1(t)$ saturates as $W_1(t) \propto 1/\sqrt{t}$ to a value set by the measured subsystem size $L_B$. 
             Relative errors are all below $10\%$, and are not shown, for $R=10^4$.
             An empirical PE is constructed every (10) 100 time steps for $L_B = (6), 8, 16, 32, 64$.
             (b) Closer view of the decay $W_1(t) \propto 1/\sqrt{t}$ (dashed line) for $R=10^3$ and an empirical PE built at each time step.
             (c) Scaling of the late-time saturation value $W_1 (t\gg 1) \propto 1/L_B$ (dashed line), imperceptible error bars~(SE). 
            }
    \label{fig:app_W1_entropies}
\end{figure}

\section{Additional entanglement entropy numerics. Ensemble mean and variance}

In this appendix, for completeness, we further test the convergence of the PE to the GHE by considering the ensemble mean and variance of the entanglement entropy (EE) of a subsystem $A_1$ of size $L_{A_1} = 1, \dots, L_A$. By being finite ($k=1,2$) moments of the distribution $P(\theta)$, these indirectly test the joint PDF of up to $L_A/2$ singular values of the $\Gamma^{A_1}_\mathbf{m}$ block of the CM $\Gamma^A_\mathbf{m}$. 
There are known exact expressions for the mean $\langle S(f) \rangle_\mathrm{GHE}\coloneq \mathbbm{E}_{\Gamma^A \sim \Upsilon^A_\mathrm{pure}} S(\psi_{A_1})$ and variance $\Delta S(f)^2_\mathrm{GHE} \coloneq \mathbbm{E}_{\Gamma^A \sim \Upsilon^A_\mathrm{pure}} \left[ S(\psi_{A_1})  - \langle S(f) \rangle_\mathrm{GHE}\right]^2$ of the EE in the GHE~\cite{bianchi_pageFGS,yu_pageFGS}, where $f \coloneq \frac{L_{A_1}}{L_A}$. These are distinct from qubit systems~\cite{pageCurve, pageVariance}, and their dependence on $L_A$ and $f$ is discussed in App.~\ref{app:EE}.

By estimating or computing exactly $\langle S(f) \rangle_\mathrm{PE}$ and $\Delta S(f)_\mathrm{PE}$, we further support the convergence of the PE to the GHE. First, we consider local dynamics [Fig.~\ref{fig:setup}(a)], sample $R$ states from the PE, and thus estimate $\langle S(f,t) \rangle_\mathrm{PE}$ and $\Delta S(f,t)_\mathrm{PE}$ at each time step $t$. The numerics show that, at late-time, $\langle S(f,t) \rangle_\mathrm{PE}$ and $\Delta S(f,t)_\mathrm{PE}$ for all $L_{A_1} = 1, \dots, L_A$ agree well with those of the GHE, up to corrections due to the finite $L_B$ [Fig.~\ref{fig:EE_fig}(a) and inset]. Then, we proceed similarly for the global dynamics, and we also numerically compute exactly $\langle S(f) \rangle_\mathrm{PE}$ and $\Delta S(f)_\mathrm{PE}$, for a limited $L_B=20$. The estimated and exact numerics support the agreement between the PE and the GHE [Fig.~\ref{fig:EE_fig}(b)].

\begin{figure}[h]
    \centering
        \includegraphics[width=8.6cm]{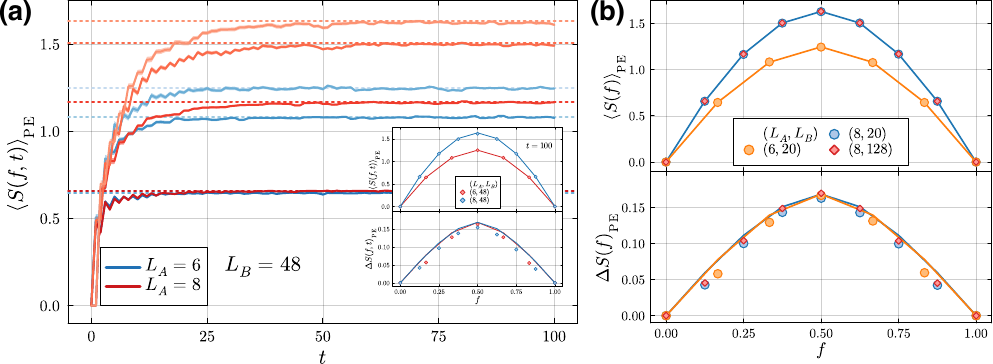}%
    \caption{%
            Ensemble mean (Gaussian Page curve) and variance of the entanglement entropy (EE).
            (a) EE mean vs time for local dynamics. The errors (SE) are shown as ribbons. We use $R=10^4$ and $N=10^2$ circuit instances. The $L_{A_1}=1,\dots,\frac{L_A}{2}$ (darker to lighter) curves reach the Gaussian Haar values (horizontal dashed lines) at late-time. Inset: EE mean and variance at $t=100$ in the PE (rhombi, imperceptible error bars) and the GHE (solid lines) vs $f\coloneq \frac{L_{A_1}}{L_A}.$
            (b) EE mean and variance vs $f$ for global circuits. 
            The empirical PE (rhombi, $R=10^5$) and the full PE (circles) data have~$N=10^2$. 
            }
    \label{fig:EE_fig}
\end{figure}

\section{Comparison of matchgate circuits to Clifford and dual unitary circuits}

		In this appendix, we highlight the key aspects that make it possible to experimentally probe deep thermalization in matchgate circuits without exponential post-selection overheads, and make a comparison to Clifford and dual unitary circuits, which are not suitable.
		
		To overcome the postselection problem using the hybrid classical-quantum methods presented in Refs.~\onlinecite{psf2, psf3}, one must be able to estimate Born probabilities and properties of specific post-measurement states through a classically efficient algorithm, i.e., taking $\mathrm{poly}(L)$ time and memory. Here we consider several different classes of circuits in turn, and assess their suitability for experimental studies of this kind. %
		
		\subsection{Matchgate circuits}
		The main advantage of matchgate circuits is that there is a regime where the projected ensemble approaches a nontrivial universal ensemble---the Gaussian Haar ensemble (GHE)---while the post-measurement states can be classically simulated. In fact, the necessary simulation task can be achieved even if these circuits with a small number [up to $\mathcal{O}(\log L)$] of non-Gaussian gates \cite{Dias2024classicalsimulation}. Thus, the PE of matchgate-dominated circuits has the twofold advantage of having an interesting structure, and could be probed experimentally without exponential postselection overheads.

		\subsection{Clifford circuits}
		Circuits composed of Clifford gates and Pauli measurements, which generate stabilizer states as output, can also be efficiently simulated classically; thus, properties of the projected ensemble can, in principle, be studied experimentally. However, in this case, the projected ensemble does not exhibit non-trivial universal phenomena, as shown by the following result.
		\begin{lemma}\label{lemma:CPE}
			The projected ensemble for a pure stabilizer state $\ket{\Psi^{AB}}$---obtained by measuring $L_B$ independent and commuting Pauli operators, supported on subsystem $B$ with $L_B$ qubits---consists of states that are pairwise equal or orthogonal. Such an ensemble cannot form a quantum state $k$-design on $A$ for any $k > 1$.
		\end{lemma}
		\noindent We prove the above in Section \ref{app:CliffordLemma}.  Note that a 1-design simply means that the reduced density matrix $\rho^A$ is maximally mixed, and is very different from a random stabilizer state, which would form a 3-design.  From this result, we expect the Clifford-dominated circuits will \textit{not} approach any form of maximally random ensemble.\\

		\subsection{Dual unitary circuits}
		
		Dual unitary circuits are a special class of one-dimensional quantum circuits for which certain quantities can be computed efficiently. While these circuits typically deeply thermalize in time $t \propto L_A$~\cite{HoChoi2022, Claeys2022}, the specific quantities that one needs to evaluate to overcome the postselection problem are not those that can be simulated classically. In fact, the task of computing Born probabilities or post-measurement states is as hard as for a general unitary circuit, because applying projectors to the measured qubits prevents the circuit from being simplified in the way that is typical for dual unitary circuits. One needs to contract a highly (both spatially and temporally volume-law) entangled two-dimensional tensor network corresponding to at least $\mathcal{O}(L_A) \times \mathcal{O}(t)$ qubits. Since this contraction requires $\exp(\min (L_A,t))$ memory and/or time for a classical computer~\cite{Markov_2008}, it is classically hard for sufficiently large $L_A$ and $t$.

		\subsection{Proof of Lemma \ref{lemma:CPE} \label{app:CliffordLemma}}
		
		Let $\ket{\Psi^{AB}}$ be the pre-measurement stabilizer state, whose stabilizer group $\mathcal{S} = \langle s_1, \ldots, s_L\rangle$ is generated by a set of $L$ commuting Pauli strings $\{s_i\}$. Without loss of generality, we consider measuring each qubit in $B$ in the $Z$ basis, with outcomes $\mathbf{m} \in \{0,1\}^{L_B}$. Because the measurement basis is Pauli, the post-measurement states $\ket{\phi_A(\mathbf{m})}$ are themselves stabilizer states \cite{Nielsen2010}. Let $\mathbf{m}$ be any bitstring that occurs with nonzero probability $p_\mathbf{m} = \braket{\Psi^{AB}|(\mathbbm{1}_A \otimes \ket{\mathbf{m}}\bra{\mathbf{m}})|\Psi^{AB}} \neq 0$. We claim that for any other outcome $\mathbf{m}'$ for which  $p_{\mathbf{m}'} \neq 0$, there exists a Pauli string $P_{\mathbf{m}', \mathbf{m}}$ such that the post-measurement states are related by
		\begin{align}
			\ket{\phi_A(\mathbf{m}')} = P_{\mathbf{m}', \mathbf{m}} \ket{\phi_A(\mathbf{m})}.
			\label{eq:CliffordPMRelation}
		\end{align}
		As a consequence, the overlaps can be written as a Pauli expectation value $\braket{\phi_A(\mathbf{m})|\phi_A(\mathbf{m}')} = \braket{\phi_A(\mathbf{m})|P_{\mathbf{m}', \mathbf{m}}|\phi_A(\mathbf{m})}$. By standard properties of stabilizer states, such expectation values can only take values in $\{-1,0,1\}$, i.e., the two states are either orthogonal or equal (up to a sign). Hence, the frame potential $\mathcal{F}^{(k)} \coloneqq \Tr[(\mathcal{E}^{(k)})^2]$ for the PE is
		\begin{align}
			\mathcal{F}^{(k)}_\mathrm{PE} = \Tr[(\mathcal{E}^{(k)}_\mathrm{PE})^2] = \sum_{\mathbf{m}, \mathbf{m}'} p_{\mathbf{m}} p_{\mathbf{m}'} |\braket{\phi_A(\mathbf{m})|\phi_A(\mathbf{m}')}|^{2k} = \Tr[(\rho^A)^2] \geq 2^{-L_A}
			\label{eq:CliffordFP}
		\end{align}
		independently of $k$, where we used $|\braket{\phi_A(\mathbf{m})|\phi_A(\mathbf{m}')}|^{2k} = |\braket{\phi_A(\mathbf{m})|\phi_A(\mathbf{m}')}|^{2}$. For the PE to form a $k$ design, we must have $\mathcal{F}^{(k)}_\mathrm{PE} \overset{!}{=} \mathcal{F}^{(k)}_\mathrm{HE}$ with $(\mathcal{F}^{(k)}_\mathrm{HE})^{-1} = {2^{L_A} + k - 1 \choose k}$ [cf. Theorem 2 in Ref.~\onlinecite{frame_potential}, with an appropriate rescaling of each $(\ket{\phi_i}\bra{\phi_i})^{\otimes t}$ by a factor of $n^{-1}$ corresponding to a uniform distribution of states], and this is inconsistent with Eq.~\eqref{eq:CliffordFP} for $k > 1$. The lemma then follows.
		
		All that remains is to prove Eq.~\eqref{eq:CliffordPMRelation}. We proceed by induction, considering the $Z$-basis measurement of each individual qubit in $B$ sequentially, from 1 to $L_B$. Let $\mathbf{m}_{[r]}$, $\mathbf{m}_{[r]}'$ denote the first $r$ bits of the measurement outcomes $\mathbf{m}$, $\mathbf{m}'$, respectively, and denote the state after these $r$ qubits have been measured as $\ket{\phi(\mathbf{m}_{[r]})}$, similarly for $\mathbf{m}'$. We hypothesize that there is a Pauli string $P^{(r)}_{\mathbf{m}, \mathbf{m}'}$ supported on $AB \backslash [r]$ such that $\ket{\phi(\mathbf{m}_{[r]}')} = P^{(r)}_{\mathbf{m}, \mathbf{m}'} \ket{\phi(\mathbf{m}_{[r]})}$, and we aim to show that an analogous relationship holds after qubit $(r+1)$ is measured. %
		We decompose $P^{(r)}_{\mathbf{m}, \mathbf{m}'} = P' \otimes T_{r+1}$ where $T_{r+1}$ is a Pauli matrix on qubit $(r+1)$, and $P'$ denotes the rest of the Pauli string. Because $P'$ commutes with the measurement on $(r+1)$, we only need to show that there exists a Pauli string $Q$ such that
		\begin{align}
			(\mathbbm{1}_{A B^{(r+1)}}\otimes \bra{m'_{r+1}}) T_{r+1} \ket{\phi(\mathbf{m}_{[r]})} =  (Q \otimes \bra{m_{r+1}}) \ket{\phi(\mathbf{m}_{[r]})}
			\label{eq:CliffordInductive}
		\end{align}
		in which case the desired result holds with $P^{(r+1)}_{\mathbf{m}, \mathbf{m}'} = Q P'$ (up to an irrelevant phase). We assume that $m'_{r+1} = m_{r+1}$; if $m'_{r+1} \neq m_{r+1}$, then we can replace $T_{r+1}$ with $X_{r+1}T_{r+1}$ and proceed with a similar argument. Moreover, we have $\bra{m_{r+1}}T_{r+1} = \pm \bra{m_{r+1}}Z_{r+1}T_{r+1}$, so the only cases we need to consider are $T_{r+1} = \mathds{1}_{r+1}$ or $X_{r+1}$. In the former case, Eq.~\eqref{eq:CliffordInductive} holds immediately with $Q = \mathds{1}$. This leaves only the latter case $T_{r+1} = X_{r+1}$. If there exists a stabilizer of $\ket{\phi(\mathbf{m}_{[r]})}$ that takes the form $Q \otimes X_{r+1}$, then we are done. The only scenario where there exists no such stabilizer is if $\ket{\phi(\mathbf{m}_{[r]})}$ is stabilized by either $\pm Z_{r+1}$ or $\pm Y_{r+1}$; in either case, $\ket{\phi(\mathbf{m}_{[r]})} = \ket{\phi(\mathbf{m}_{[r+1]})}\otimes \ket{\chi_{r+1}}$ and so the measurement does not alter the state of the remaining unmeasured qubits and we can set $Q = \mathds{1}$. This same logic also holds for the base case $r = 0$, and Eq.~\eqref{eq:CliffordPMRelation} follows by induction. \hfill $\square$

\end{onecolumngrid}

\end{document}